\documentclass[12pt,preprint]{aastex}

\shorttitle{Gas In/Out-Flow Rates in 7 NUGA Galaxies}
\shortauthors{Haan et al.}

\begin{document}
   \title{Dynamical Evolution of AGN Host Galaxies - Gas In/Out-Flow Rates in 7 NUGA Galaxies}

   \author{Sebastian Haan}
   \affil{Max-Planck-Institut f\"ur Astronomie (MPIA),
              K\"onigstuhl 17, 69117 Heidelberg, Germany}
   \email{haan@mpia.de}
        
   \author{Eva Schinnerer}
   \affil{Max-Planck-Institut f\"ur Astronomie (MPIA),
              K\"onigstuhl 17, 69117 Heidelberg, Germany}
   \email{schinner@mpia.de}

   \author{Eric Emsellem}
   \affil{Universit\'e de Lyon, Universit\'e Lyon~1, Observatoire de Lyon, 9 avenue  Charles Andr\'e, Saint-Genis Laval, F-69230, France ; CNRS, UMR 5574, Centre de Recherche Astrophysique de Lyon ; Ecole Normale Sup\'erieure de Lyon}

   \email{emsellem@obs.univ‐lyon1.fr}

   \author{Santiago Garc{\'{\i}}a-Burillo}
   \affil{Observatorio Astronomico Nacional (OAN)-Observatorio de Madrid,
	     Alfonso XII, 3, 28014-Madrid, Spain}
   \email{burillo@oan.es}

   \author{Francoise Combes}
   \affil{Observatoire de Paris, DEMIRM, 61 Av. de l Observatoire,
	     75914-Paris, France}
   \email{francoise.combes@obspm.fr}
	  
   \author{Carole G. Mundell}
   \affil{Astrophysics Research Institute, Liverpool John Moores University, Twelve Quays House, Egerton Wharf, Birkenhead, CH41 1LD, UK}
   \email{cgm@astro.livjm.ac.uk}

  \author{Hans-Walter Rix}
   \affil{Max-Planck-Institut f\"ur Astronomie (MPIA),
              K\"onigstuhl 17, 69117 Heidelberg, Germany}
   \email{rix@mpia.de}

\begin{abstract}
To examine the role of the host galaxy structure in fueling nuclear activity, we estimated gas flow rates from several kpc down to the inner few 10 pc for seven nearby spiral galaxies, selected from the NUGA sample (NUclei of GAlaxies). We calculated gravitational torques from near-IR images and determined gas in/out-flow rates as a function of radius and location within the galactic disks, based on high angular resolution interferometric observations of molecular (CO using PdBI) and atomic (HI using the VLA) gas. The results are compared with kinematic evidence for radial gas flows and the dynamical state of the galaxies (via resonances) derived from several different methods. We show that gravitational torques are very efficient at transporting gas from the outer disk all the way into the galaxies centers at $\sim$100~pc; previously assumed dynamical barriers to gas transport, such as the Corotation Resonance of stellar bars, seem to be overcome by gravitational torque induced gas flows from other non-axisymmmetric structures. The resulting rates of gas mass inflow range from 0.01 to 50~M$_{\odot}$~yr$^{-1}$ and are larger for the galaxy center than for the outer disk. Our gas flow maps show the action of nested bars within larger bars for 3 galaxies. Non-circular streaming motions found in the kinematic maps are larger in the center than in the outer disk and appear to correlate only loosely with the in/out-flow rates as a function of radius. We demonstrate that spiral gas disks are very dynamic systems that undergo strong radial evolution on timescales of a few rotation periods (e.g. $5\cdot10^8$~yrs at a radius of 5~kpc), due to the effectiveness of gravitational torques in redistributing the cold galactic gas.
\end{abstract}


%

\section{Introduction}
\label{sec:intro}
The study of gas inflow rates has long been recognized to be important for understanding the secular evolution of galaxies, star formation, and nuclear activity. 
Generally gas inflow can be caused by two different dynamical mechanisms, gravitational and hydrodynamic mechanisms: galaxy interactions and asymmetries in galactic potentials remove angular momentum through torques, while hydrodynamic mechanisms, such as turbulence in the ISM, remove angular momentum through gas dynamical effects (e.g. viscous torques or shocks). Here we focus on the first mechanism with emphasis on the gas response to gravitational torques exerted by the stellar potential. 
It is well established that asymmetries in galactic potentials, such as large-scale bars, transport gas very efficiently inward \citep[e.g.,][]{Mun99, Com03, Jog04}, and that bars can fuel powerful starbursts \citep[e.g.,][]{Jog05, Sch06, Hun08}. Further, the inflow of gas from the outer regions of a galaxy is necessary to maintain nearly constant star formation in the inner disks of spirals \citep{Bli96} and to form (and refuel) an Active Galactic Nuclei (AGN) accretion disk around a central black hole \citep{Ree84, Shl90}. \par

However, searches for low-luminosity AGN fueling mechanisms have not found any strong correlation between the presence of morphological features, e.g. bars, and AGN activity \citep[see for a review][]{Com01, Shl03, Kna05, Jog06}. High spatial resolution studies in the near infrared (NIR) have shown that most galaxies are barred (60-80\%), almost irrespective of their nuclear activity, and that a significant number of Seyfert host galaxies exist in which no bar can be distinguished at practically all length scales \citep{Kna00, Lai02}. 
One explanation might be that smaller-scale phenomena are responsible for AGN fueling. Thus, only observations with very high spatial resolution provide the possibility to find correlations between AGN fueling and dynamical perturbations. Martini \& Pogge (\citeyear{Mar99}) suggested that nuclear spiral dust lanes may be the channels by which gas from the host galaxy disks is being fed into the central engines. But also a high spatial resolution study with the Hubble Space Telescope (HST) of the circumnuclear region (on 100~pc scales) has still not found a significant difference between AGN and control samples, at least not for late-type galaxies \citep{Mar03, Sim07}.
Another possibility is that a hierarchy of mechanisms combine to transport the gas from the outer disk (kpc scales) down to the center (pc scales). As these various mechanisms are working at different spatial scales, also the mass accretion involves different time scales \citep{Shl90, Com03, Wad04}. This time-scale conspiracy could explain the lack of any correlation between the presence of non-axisymmetric perturbations and the onset of activity \citep{Wad04,Gar05}. Recent spectroscopic studies of the ionized gas in the central kiloparsec seem to confirm this picture by showing evidence for kinematic perturbations linked with Seyfert galaxies at small radii \citep{Dum07}. For larger scales ($>1$~kpc) first results of a detailed study of HI gas properties for active and non-active disk galaxies indicate possible relationships between Seyfert activity and HI properties \citep{Mun07, Haa08}.\par

As the neutral gas is distributed in a rotating disk and extends out to kpc scales, the problem of radial gas transport is basically a problem of angular momentum removal from the gas. To transfer angular momentum, effective torques are required. In particular, gravitational torques, exerted by the stellar and effective dark matter potential are expected to play a major role \citep[see for an overview][]{Lyn72}. In addition, viscous and magnetic torques might contribute significantly. For example, viscous torques are expected to be very efficient in high gas density regions such as the centers of galaxies \citep{Lyn69, Gar05}. However, normal viscosity is not efficient, due to the very low density of the gas. Even with macroturbulent viscosity, the time-scales are longer than the Hubble time at large radii, and could be effective only inside the central 1~kpc \citep{Lyn72}. The role of magnetic torques is more uncertain since the interplay between magnetic fields and gas is still not very well understood. Although magnetic fields seem to pervade the interstellar gas in spiral galaxies \citep{Bec99, Bec04}, it is more likely that magnetic fields follow the gas flow rather than acting as incitement. \par

The quest for a quantitative estimation of radial gas flows from observations has been pursued by different methods: One approach is to use the gas kinematics from radio-interferometric observations of moderately inclined spiral galaxies and a decomposition of the velocity field into Fourier components whose radial variations are used to search for evidence of radial gas flows \citep[e.g.][]{Sch97, Sch99, Jog02, Fat04, Won04,Kra06,Tra08}. This method has also been successfully applied using optical spectrography in order to search for perturbations in the stellar dynamics \citep[][]{Dum07, Dum07t, Ems06}. Using a slightly different approach, \cite{Boo07} derived gravitational torques by fitting a model based on analytical solutions for gas particle orbits for the barred spiral galaxy NGC~4569. \cite{Spe07} developed a bisymmetric flow model for non-circular motions that can be fitted to data (applied to the galaxy NGC~2976) by a generalization of the technique developed by \cite{Bar03}. However, the conversion of the model derived from the velocity field into radial gas flows depends strongly on the chosen model for the gas orbits. As the gas orbits are generally not known, it is impossible to recover the velocity structure from the line-of-sight component of motion alone without further assumptions \citep[e.g. using isophote fits on complementary optical images,][]{Won04}. \par

Another approach for measuring inflow velocities is to derive the gravitational potential on the basis of the observed near-infrared (NIR) light distribution of a galaxy and then to compute the gravitational torques exerted by the stellar potential on the gas \citep{Gar93, Qui95, Gar05}.

In combination with observed rotation curves this method does not depend necessarily on assumed mass-to-light ratios (hereafter M/L ratio) which are derived, e.g. by population synthesis techniques, and are inaccurate for the outer disk where dark matter is expected to have a significant contribution. Instead, the azimuthal averaged calculated velocities from the derivatives of the gravitational potential can be scaled with the actual observed rotation curve and hence allow one to obtain a more realistic scaling of the gravitational potential (assuming a constant azimuthal M/L ratio). \par

In fact, gravitational torques arise from non-axisymmetric components of the gravitational potential, such as stellar bars, spirals, and oval distributions. The effect of large-scale bars has been studied in detail by simulations which have successfully shown that bars are very efficient for transporting the gas towards the center \citep[e.g.][]{Spa87, Ath92, Shl93, Hel94, Ath03}. As these mechanisms may work on different spatial scales, secondary bars within large-scale bars \citep[e.g.][]{Shl89, Friedli, Mac00, Hel01, Shl02, Eng04}, gaseous spiral density waves \citep[e.g.][]{Eng00, Mac02, Mac04a, Mac04b}, m = 1 perturbations \citep[e.g.][]{Shu, Jun96, Gar00} and nuclear warps \citep{Sch00} have been suggested to transport the gas into the center. \par

In order to distinguish models for nuclear fueling, observations with high angular and velocity resolution are required. Therefore, the IRAM key project NUclei of GAlaxies \citep[NUGA; see][]{Gar03} was established - a spectroscopic imaging survey of gas in the centers of nearby low luminosity AGN. As most of the gas in galaxy nuclei is in the molecular phase, the survey used millimeter CO lines to conduct a detailed mapping of molecular gas dynamics at high-resolution ($\sim0.5\arcsec$) in the central kiloparsec of AGN hosts. To provide a more complete view of gas dynamics from the outer disk to the center, the HI-NUGA project has been initiated: a spectroscopic imaging survey of the atomic gas distribution and kinematics over the entire galaxy disks ($\sim$25~kpc) for 16 galaxies of the NUGA sample \citep{Haa08}. Results of this study indicate that the type of nuclear activity (Seyfert/LINER) is linked to the gas distribution in the outer gaseous disk (at several~kpc scales), suggesting a time evolution of AGN activity together with the redistribution of the neutral gas due to the non-axisymmetric potential.    
As the dominant phase of the neutral ISM transitions from atomic to molecular gas towards the center of galaxies, the combined HI and CO NUGA observations cover all scale lengths from the outer disk at $> 10$~kpc to the center at $\sim 100$~pc. \par

In this paper we are studying in detail the gas dynamics and the gas response to the gravitational potential for 7 well-chosen targets from the HI-NUGA sample. We present a novel approach to derive gas inflow rates via a combination of gravitational torque computation based on the stellar distribution and a kinematic analysis of the observed gaseous velocity fields.
To derive gravitational torques, high spatially resolved maps of the stellar and gaseous distribution are required. Therefore we have obtained for our 7 galaxies VLA HI data with $\sim7 \arcsec$ spatial and $\sim$10~km~s$^{-1}$ velocity resolution. Our sample and the observational setups (HI-, CO- line emission, and NIR data) are described in \S \ref{sec:obs}. The dynamical models, the geometric disk parameters, and the amount of non-circular motions are derived in \S \ref{sec:kin} from our observed velocity fields using the atomic and molecular gas kinematics. We have computed gravitational torques and estimated their efficiency to transport the gas over galaxy disks on the basis of NIR images (Spitzer, HST) and a combination of our HI data and NUGA CO data with $\sim 0.5 \arcsec$ spatial and $\sim$5~km~s$^{-1}$ velocity resolution. The methods used for this study and its results are described in \S \ref{sec:torque}. Finally, we compared in \S \ref{sec:dis} the total amount of non-circular motions from our observed velocity fields to the non-circular motions that are induced by the gravitational torques and discuss our findings in the context of AGN fueling mechanisms. A summary is given in \S \ref{sec:sum}.

\section{Sample Description and Observations}
\label{sec:obs}
For our study we have used observations of the neutral gas component as traced by HI and CO, as well as the stellar component as described in the following.

\subsection{Sample Description}
\label{subsec:obs_sample}
For a detailed study of gas inflow rates, we selected 7 galaxies from our HI-NUGA sample \citep{Haa08}. The selection of the targets is based on the presence of bright HI emission as well as a fairly regular HI morphology (i.e. no highly disturbed gas disks). All the galaxies are spiral galaxies ranging in Hubble type from Sa to Sc and are barred, with the exeception of NGC~7217. The distance of our sample targets ranges from 4 to 24~Mpc with a mean distance of $\sim$13~Mpc. A complete overview is given in Tab.~\ref{tab_overview}. The galaxies host various kinds of nuclear activity\footnote{The nuclear classification is adopted from \cite{Haa08} and based on optical emission line ratios following the method of \cite{Kew06}}: LINERs (4 galaxies), Seyferts (2 galaxies), and starburst (1 galaxy). 

\subsection{Atomic Gas Data}
\label{subsec:obs_HI}
To trace the dynamics of the outer disk we observed the atomic gas distribution and kinematics in the HI emission line at 21cm using the NRAO Very Large Array (VLA). Previous data obtained with the VLA in C and D array configuration ($\sim$20$\arcsec$ resolution) were already presented by \cite{Haa08}. For the 7 galaxies analyzed here we combined these data with newly obtained VLA B-array data resulting in a final resolution of $\sim 7 \arcsec$. Two galaxies (NGC~3627, NGC~4736) were already observed as part of the THINGS project \citep{Wal08}. The data reduction was performed using the Astronomical Image Processing System \citep[AIPS;][]{greisen03} following the data reduction scheme of the previous CD array data \citep{Haa08}.
Flux calibrator measurements were performed at the beginning and at the end of each observation cycle. The phase calibrator was observed before and after each source cycle with a maximum angular distance between source and phase calibrator of 12$\degr$.
The data have an average on-source integration time of 8.5h (B-array), 2.7h (C-array), and 2.8h (D-array). The correlator spectral setup used was set to line mode 4 with Hanning smoothing and 64 channels per 1.5625 MHz channel width per frequency band providing a frequency resolution of 24.414 kHz/channel ($\sim$ 5.2~km~s$^{-1}$). Calibration solutions were first derived for the continuum data-set (inner 3/4 of the spectral band width) and then transferred to the line data. The bandpass solutions were determined from the phase calibrator measurements to account for channel to channel variations in phase and amplitude. The B and CD data have been combined. The CLEANing parameters are adjusted to the new combined BCD data, as described in the following. The cellsize of each grid was set to 1.3$\arcsec$/pixel with a field of view (FOV) of $\sim 22\arcmin$. We produced the CLEANed data cubes using robust weighted imaging with a velocity resolution of $\sim10.4$~km~s$^{-1}$ and an average angular resolution of $\sim 7\arcsec$.  To find the best compromise between angular resolution and RMS, several robust weighting parameters were tested and a robust parameter of 0.3 was selected. The RMS values and beam sizes are listed in Table~\ref{tab_obs} with an average achieved RMS value of 0.32 mJy beam$^{-1}$. The RMS flux sensitivity of 0.4 mJy/beam/channel corresponds to a 3-$\sigma$ detection limit of $\sim 0.55 \times10^{19}$ cm$^{-2}$ column density for the combined BCD array data.
To separate real emission from noise, we produced masks by taking only into account those regions which show emission above a set level (3$\sigma$) in cubes that have been convolved to 30$\arcsec$ resolution (task CONVL). Using these masks, we blank areas that contain noise in our robust weighted data cubes (task BLANK).
\par
The subsequent analysis has been done with the Groningen Image Processing SYstem \citep[GIPSY;][]{Hulst}. The channel maps were combined to produce zeroth (intensity map), first (velocity field) and second (dispersion map) moments of the line profiles using the task MOMENT. The RMS values have been measured in two regions where no line signal was apparent and averaged over all channels of the cubes. A flux cut-off of three times the channel-averaged RMS value was used for the moment maps. The velocity-integrated HI intensity maps are presented in Figure \ref{fig_kin}.

\subsection{Molecular gas data}
\label{subsec:obs_CO}
The spectroscopic imaging of CO lines provides information about the distribution and kinematics of the molecular gas. 
The molecular gas has been observed in the transition of the J = 1$-$0 and J = 2$-$1 lines of $^{12}$CO with maximum angular ($\sim 0.5\arcsec$) and spectral resolution (3-6~km~s$^{-1}$) using the IRAM Plateau de Bure mm-interferometer (PdBI) as part of the NUGA project and the larger NUGA supersample \citep{Gar03}. The primary beam size is $\sim$42$\arcsec$ ($\sim$21$\arcsec$) in all the CO 1$-$0 (2$-$1) line observations. Only for NGC~5248 no PdBI data was available. Therefore we used for NGC~5248 the publicly available BIMA-SONG data \citep{Hel03} which was observed in the CO (1-0) line with an angular resolution of 6.1$\arcsec$ and spectral resolution of 10~km~s$^{-1}$ with the BIMA mm-interferometer. An overview of the observational parameters for the CO data is presented in Tab.~\ref{tab_COobs}.

\subsection{Near Infrared Images}
\label{subsec:obs_NIR}
Archival near infrared images (NIR) from Spitzer, HST and ground-based telescopes have been used to estimate the stellar mass distribution in our galaxies. An overview of all NIR images used is presented in Tab.~\ref{tab_NIR}. Foreground stars have been removed using the software tool NFIGI (Baillard et al., in prep.) and, in case additional cleaning was necessary, by hand. In addition, background is subtracted from the NIR images by calculating the mean value in the outer region of the image outside the galaxy disk to ensure that no light contribution from the galaxy itself is subtracted. Other methods for background subtraction have been tested (e.g. fitting the radial luminosity profile to find constant background value), but could not be applied successfully to all NIR images because of limited FOVs. After background subtraction all values below zero, which might be caused by small overestimation of the background value (case only for outer regions of image), were set to zero. 

\section{Kinematic Modeling}
\label{sec:kin}
To search for kinematic evidence for non-circular motions and radial gas flows, we have performed a detailed kinematic analysis of our CO and HI velocity fields. In \S \ref{subsec:kin_obs} we first estimate the basic disk orientation parameters that are required for our gravitational torque calculations. Then, we model non-circular motions and search for evidence of gas inflow or outflow by employing a Fourier analysis of the velocity fields as described in \S \ref{subsec:kin_model}. The results of this study are presented in \S \ref{subsec:kin_results} and indicate, e.g., a higher ratio of non-circular to circular gas motions for the central (CO) than for the outer gaseous disk (HI).

\subsection{HI and CO kinematics}
\label{subsec:kin_obs}
\begin{figure}[ht]
\includegraphics[scale=0.55]{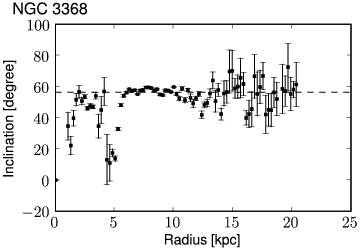}
\includegraphics[scale=0.55]{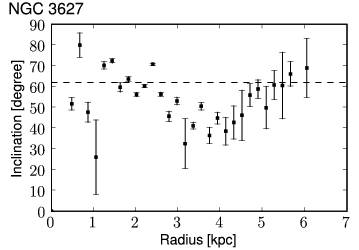}
\includegraphics[scale=0.55]{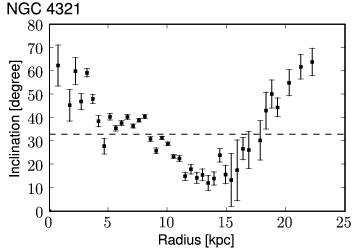}
\includegraphics[scale=0.55]{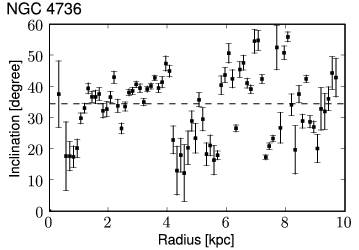}
\includegraphics[scale=0.55]{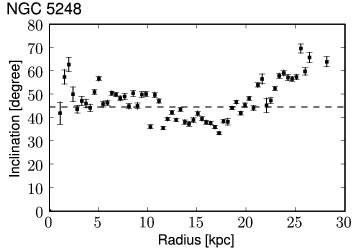}
\includegraphics[scale=0.55]{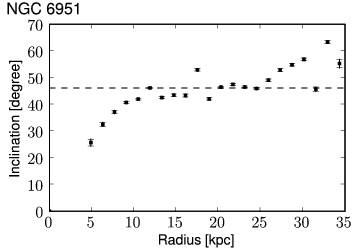}
\includegraphics[scale=0.55]{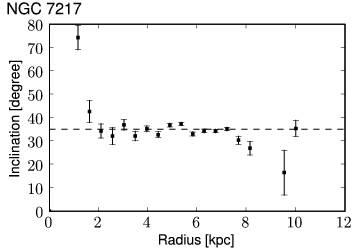}
\caption{ Overview of the inclination angle parameters as a function of radius derived from fitting the HI velocity field. The dashed line indicates the calculated mean inclination angle.}
\label{fig_incl}
\end{figure}

In preparation for our analysis of radial gas flows, we derived the following disk parameters by fitting tilted rings to the CO and HI velocity fields using the task ROTCUR within GIPSY:
\begin{itemize}
\item The dynamical center and its offset from the optical center (taken from Hyperleda).
\item The systemic velocity $v_{sys}$ in km~s$^{-1}$.
\item The position angle (PA) in degrees, defined as the angle between the north-direction on the sky and the receding half of the major axis of the galaxy in anti-clockwise direction.
\item The inclination ($i$) in degrees.
\end{itemize}
The kinematic parameters have been derived in an iterative way \citep{Beg89} as described in our analysis for the entire HI-NUGA sample \citep{Haa08}. The parameters were assumed to be the same at all radii, except for the circular velocity. We weighted the obtained value in each ring with its standard deviations, in order to derive the mean parameter. For the fit we excluded data points within an angle of 20$\degr$ of the minor axis. The widths of the radii were set to 0.4$\arcsec$ (6$\arcsec$) which corresponds roughly to the angular resolution of our CO (HI) data. No radial velocity component was fitted as disk parameter. We have applied the same systemic velocity, inclination, and PA derived from our HI velocity field for the CO analysis as we assume that these disk orientation parameters are constant over the entire galaxy disk. We found no evidence for significantly warped disks in our sample using the derived inclination angle as a function of radius (see Fig.~\ref{fig_incl}). For the estimation of the dynamical center position we have taken into account only data of the inner third of the HI gas disk to exclude possible shifts of the kinematic center due to spiral arms and disturbances in the outer HI disk. In addition we derived the center position from the CO kinematics for most galaxies of our sample. However, our derived values for the kinematic center show only small offsets from the photometric center with an average offset of $\sim$1.5$\arcsec$ which is similar to typical errors expected from photometric center fitting. An overview of all derived parameters for each galaxy is presented in Tab.~\ref{tab_par}. Finally, the rotation velocities were obtained for the atomic and molecular gas with fixed inclination, position angle, center and systemic velocity.

\subsection{Description of kinematic modeling of non-circular motions}
\label{subsec:kin_model}
In this study we attempt to detect non-circular motions and subsequently evidence for radial gas flow directly from the observed gas kinematics. In principle, one would expect that non-circular motions are caused by the non-axisymmetric part of the gravitational potential. However in practice, other physical effects such as gas viscosity and shocks in the gas may have significant contributions to any observed non-circular motion. Furthermore, interactions with galaxy companions as well as AGN and star formation feedback might work as additional source for non-circular motions. To estimate the contribution from these effects, we compare the measurement of the non-circular motions with the results obtained via our gravitational torque analysis (see \S \ref{subsec:dis_comp}). To obtain a fair estimate of the circular and non-circular part of our gas velocity fields, we have used the GIPSY task RESWRI which performs a harmonic expansion of the velocity fields. This expansion is made by first fitting a tilted-ring model to the velocity field of the gas disk and subsequently decomposing the velocity field along each ring into its harmonic terms. Since the data points may not be uniformly distributed in azimuth, RESWRI performs a least-square fitting (singular value decomposition) rather than a direct Fourier expansion. In practice, after the convergence of the ROTCUR part, RESWRI makes a harmonic expansion of the line-of-sight velocity $ v_{los}$ along each ring,
\begin{equation}
 v_{los}(r) = v_{sys}(r)+\sum_{n=1}^k \left[  c_n(r) \cos(n\psi)  + s_n(r) \sin(n\psi) \right] ,
\end{equation} 
where $k$ is the order of the fit, $r$ the radial distance from the center, $\psi\def \theta-\theta_{obs}$ the azimuthal angle, measured from the receding side of the line of nodes, and $v_{sys}$ the systemic velocity  as 0$^{th}$ harmonic component. The coefficients $c_n$ and $s_n$ are determined by making a least-squares-fit to the data points up to third order, which requires three sine terms ($s_1, s_2, s_3$) and three cosine terms ($c_1, c_2, c_3$).
Since the line-of-sight velocity is given in the most general case of a velocity field as
\begin{equation}
v_{los}(r) = v_{sys}(r)+v_{\theta}(r) \cos(\psi) \sin(i) + v_R(r) \sin(\psi) \sin(i)
\end{equation} 
where $v_{\theta}$ and $v_R$ are the circular and radial components of the velocity field, respectively. Hencefore, 
$c_1=v_{\theta}(r) sin(i)$ reflects the observed circular velocity, whereas all other terms are contributions to non-circular motions \citep[see for a detailed discussion of the harmonic terms][]{Sch97, Sch99}. \par
To estimate the total amount of non-circular motions we calculate the quadratically added amplitude of all non-circular harmonic components $v_{nc}$ up to the order of $N=3$ \citep[see also][]{Tra08}:
\begin{equation}
v_{nc}(r) = \sqrt{s_1^2(r) + c_2^2(r) + s_2^2(r) + c_3^2(r) + s_3^2(r)}. 
\end{equation} 
which is basically the vector sum of all non-circular velocity contributions.
The fraction $v_{nc}/v_{rot}$ shows how the contribution of non-circular motions to the total velocity may vary with radius.  
At the corotation radii ($R_{CR}$) of a bar or spiral the $s_3$-terms become dominant over the  $s_1$-terms \citep{Can97}.

\subsection{Results of our kinematic analysis}
\label{subsec:kin_results}
For the outer disk (probed by our HI gas kinematics), most of the galaxies in our sample exhibit fairly regular velocity fields that are dominated by circular motions (see Fig.~\ref{fig_kin}). The ratio of non-circular motions to rotational velocity $v_{nc}/v_{rot}$ averaged over the radius for HI is typically in the range of 0.05-0.09, except for NGC~3627 and NGC~4736 which exhibit a slightly larger ratio of 0.12 and 0.14, respectively. For NGC~3627, that can be explained by the fact that it belongs to the Leo Triplet and non-circular motions are very likely induced by the past encounter with NGC~3628 \citep{Zha93}. Although NGC~4736 exhibits a one-arm spiral in the outer HI disk which might be a hint for a previous interaction, no optical companion is obvious. The mean values of $v_{nc}/v_{rot}$ are derived by weighting the data points with their standard deviations and are presented together with the standard deviation of the mean value in Tab.~\ref{tab_kin}. Although the mean ratios $v_{nc}/v_{rot}$ are quite small, the ratio of non-circular motions in HI is increasing to 40\% as a function of radius, primarily in the inner- and outer-most region of the HI disk. For the center at $\lesssim$1~kpc, a larger fraction of non-circular motions is derived from our CO velocity fields with typical values of (7-14)\% for NGC~7217, NGC~5248, NGC~3368, NGC~6951 and very large fractions of (22, 33, 83)\% for NGC~4736, NGC~4321, and NGC~3627, respectively. In particular, for NGC~3627 non-circular motions seem to dominate the velocity field not only in the center, but also have significant contributions in the outer disk. \par
          
In Fig.~\ref{fig_harmonics} we plot the rotation curve $v_{rot}$ and the fraction of non-circular motion to circular motion $v_{nc}/v_{c_1}$ as a function of radius for each galaxy.
The total power of non-circular motions $v_{nc}$ is typically larger for the central (CO kinematics) than for the outer gaseous disk (HI) with mean values of 26\% and 9\%, respectively. A comparison of non-circular motions and radial flow directions obtained from our kinematics and the ones on the basis of gravitational torques is performed in \S ~\ref{subsec:dis_comp}. Further, we compare in \S \ref{subsec:dis_comp} candidate radial gas flow regions derived from our velocity fields with the inflow/outflow rates due to gravity torques.
In fact, all studies that have tried to determine an actual inflow/outflow on the basis of the harmonic decomposition method alone were not very successful so far. The main reason for that is that elliptical streaming in a bar or spiral potential seems to be the dominant contributor to noncircular motions \citep[see][for more details]{Won04}. To derive net inflow/outflow a phase shift between the gas and the gravitational potential has to be present, which is the underlying concept for our gravity torque study in the next section.  

\section{Gravitational torques}
\label{sec:torque}
In this section we focus on the study of gravitational torques exerted by the stellar potential on the gaseous disk. 
By definition, a torque is a vector \boldmath $\tau$, which points along the rotation axis it would tend to cause. It can be described mathematically as the cross product $\tau = r \times F$  where $r$ is the particle's position vector relative to the fulcrum and $F$ is the force acting on the particles, or, more generally, as the rate of change of angular momentum, $\tau=$\unboldmath$\mathrm{d}$\boldmath$L$\unboldmath$/\mathrm{d}t$ where \boldmath $L$ \unboldmath is the angular momentum vector and $t$ stands for time. 
The efficiency with which gravitational torques drain the angular momentum of the gas depends first on the strengths of non-axisymmetric components such as bars and oval distributions, but also, on the existence of significant phase shifts between the gaseous and stellar distributions \citep{Gar05}. To calculate these phase shifts high spatial resolution images of the gas and stars are necessary.
The method of mapping the gravitational potential and torques in a non-axisymmetric galaxy is based on its appearance in NIR images using a Fourier transform method \citep{Gar93, Qui94, Gar05}. These gravity torque maps are then utilized to determine the transport of angular momentum, gas inflow rates, and the efficiency of the feeding mechanism for the central black hole. For the computation of the potential and torques we have developed a software toolkit (PyPot) which makes use of the programming language Python and its associated software packages (e.g. Scipy).  \par

The major fraction of the stellar mass of spiral galaxies is made up by old stars such as cool giants and dwarfs rather than from hot young stars which are bright and blue. As near infrared images detect light primarily from these old stars due to their SED, they better trace the mass distribution of a galaxy in comparison to optical images which are biased by the contribution of bright young stars \citep{Aar77, Qui94}.
Furthermore NIR images are less affected from dust-extinction than optical ones and can reveal bars that are not easily observed in the optical \citep[e.g.][]{Qui94}.

\subsection{Method description}
\label{subsec:torque_desc}
A detailed description of estimating gravitational torques in spiral galaxies is given by \cite{Gar05}. We used a similar method to derive gravitational torques which is briefly described below and consists of the following steps:
\begin{enumerate}
	\item Evaluation of the stellar potential using high-resolution NIR images from the Spitzer telescope, ground-based telescopes, and HST. To obtain the total gravitational potential we scale the stellar potential with mass-to-light ratios which are estimated by fitting the gaseous rotation curves derived from our HI and CO observations (see \S \ref{subsec:kin_obs}). This implicitly assumes that the stellar disks are nearly 'maximal'.
	\item Computation of gravitational forces and gravity torques based on the stellar potential. 
        \item Weighting of the torque field with the gas column density in order to link the derived torque field to angular momentum variations.  
	\item Estimation of the gas flows induced by these angular momentum variations using azimuthal averages of the torques at each radius.
  \item Finally, the time-scales and gas masses associated with inflow/outflow are derived by estimating the average fraction of angular momentum transferred in one rotation. 
\end{enumerate}
As the average fraction of gas to dynamical mass for our sample is $\sim 5\%$, the total mass budget is expected to be dominated by the stellar contribution, so that we will neglect self-gravity of the gas.

\subsubsection{Computation of the gravitational potential}
The gravitational potential is computed on a cartesian grid based on the NIR images used as tracers for the stellar potential \citep[following the method of][]{Gar93}. First we deproject the NIR and gas images using the inclination $i$ and position angle PA derived from our kinematic analysis (see \S \ref{subsec:obs_NIR}). An estimation of systematic errors of our calculations due to possible parameter errors will be presented in \S \ref{subsec:torque_error}.
Since a bulge would be artificially elongated by the deprojection, the bulge component has to be excluded from the deprojection. Thus, a bulge model is created first using results from fitting a de Vaucouleurs and exponential function to the radial stellar profile representing the bulge and disk brightness profile, respectively. The bulge model is assumed to be spherical. Then the bulge component is subtracted from the NIR image. After the deprojection of the galaxy (without bulge), the bulge model (with the same scaling as before) is added again to the deprojected galaxy disk. 
The deprojected NIR and gas images are shown in Fig.~\ref{fig_torq_dpj}.

The gravitational potential of the galaxy, 
\begin{equation}
\Phi(\mathbf{x})=-G \int \frac{\rho(\mathbf{x'})d^3 \mathbf{x'} }{| \mathbf{x} -\mathbf{x'} |},
\end{equation} 

can be written as a convolution of the 3D mass density $\rho(x,y,z)$ and the
function $g(r) = 1/r$ (the gravitational potential of a point-like source).
\begin{equation}
\Phi(x,y,z) = -G \cdot \rho(x,y,z) \otimes g(r) 
\end{equation}
where $G$ is the gravitational constant.
We use a smoothing length $\epsilon$, to avoid singularities for distances
close to or equal to 0. 
This results in a slight change of the convolution kernel to:
\begin{equation}
g(r, \epsilon) = \frac{1}{\sqrt{r^2 + \epsilon^2}}
\end{equation}
and we find the normal $1/r$ function when $\epsilon$ goes to 0.
Further, the thickness of the stellar component of galactic disks is not negligible. 
Observations have shown that the vertical scale height of stellar disks is roughly constant as a function of radius \citep{Kru82a, Kru82b, Wai89, Bar92}.
To take this into account we assume a non-infinitesimally thin disk and use a model for the vertical distribution  $h(z)$, namely an isothermal plane with 
\begin{equation}
h(z)=\frac{1}{2}\;h\;sech^2(z/a),
\end{equation} 
and a constant scale height $a$, equal to $\sim 1/12$ of the radial scale length of the galaxy disk.
In that case, the mass density distribution is written as:
\begin{eqnarray}
\rho(x,y,z) & = & \Sigma(x,y) \times h(z) \\
\mbox{with} \int_{-\infty}^{\infty} h(z) & = & 1
\end{eqnarray}

To compute the potential in the equatorial plane, we integrate the contributions from all heights.
This results in a convolution kernel function which can be written as:
\begin{equation}
g(r, \epsilon) = \int_{-\infty}^{\infty} \frac{h(z)}{\sqrt{r^2 + \epsilon^2 + z^2}} \,\mbox{d}z
\end{equation}
Hence we still can write the potential in a simple form, at least in the plane of the disk:
\begin{equation}
\Phi(x,y,z=0) = -G \cdot \Sigma(x,y) \otimes g(r, \epsilon) 
\end{equation}
In practice, given the vertical thickening function $h(z)$ and the smoothing length $\epsilon$,
we tabulate the function $g(r, \epsilon)$ numerically at the beginning of our code. The convolution is then performed using Fast Fourier Transforms. Note that in order to avoid edge effects, we extend the grid to a factor of $\sim$4 times larger than the initial grid.\par

After that, the potential has to be factorized with the appropriate mass-to-light ratio which is calculated via comparing the observed rotation curve and the rotation velocities derived from the potential,
\begin{equation}
f_{scal}(R) \equiv \frac{v_{model}(R)}{v_{rot}(R)}
\label{eq_scaling}
\end{equation}
where $v_{model}$ respresents the circular velocity derived from the stellar potential and $v_{rot}$ the rotation curve velocity of our combined and interpolated CO and HI rotation curve at the radius $R$. 
Then, we employed a fit on $f_{scal}(R)$ as function of radius using a decomposition into disk, bulge, and dark matter component. The derived scaled circular velocities from the gravitational potential match very well the observed circular velocities within $\pm 5\%$ along the radial axis.

\subsubsection{Computation of gravitational forces and torques}
\begin{figure}[ht]
\begin{center}
\includegraphics[scale=0.7]{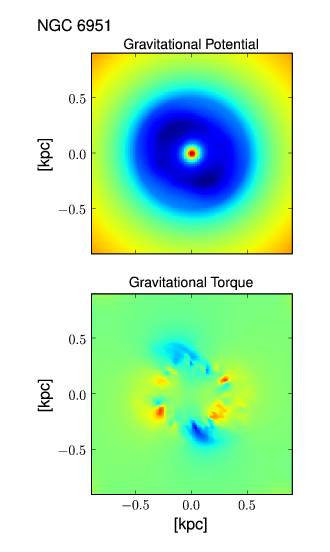}
\includegraphics[scale=0.7]{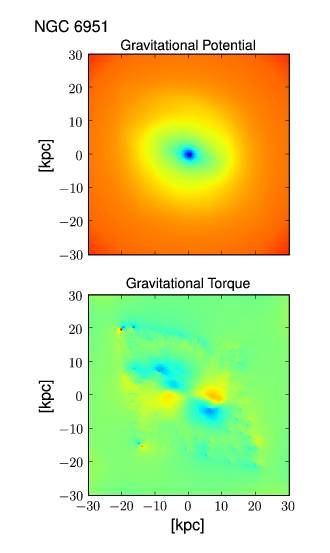}
\caption{ Example of the gravitational potential and torque (without weighting with the gas component) for the center (left panels) and the large scale (right panels) for NGC~6951.}
\label{fig_torque_pot}
\end{center}
\end{figure}

At first, the forces per unit mass in $x$ and $y$ direction $F_{x,y}$ for our derived gravitational potential are calculated at each pixel:
\begin{equation}
F_{x,y}(x,y) = -\nabla_{x,y}\Phi(x,y)
\end{equation} 
The gravity torques (per unit mass) \boldmath $\tau$ are defined as:
\begin{equation}
\tau=r \, \times \, F,
\end{equation}
\unboldmath
and calculated in the plane:
\begin{equation}
\tau(x,y)=xF_y - yF_x.
\end{equation}  
The positive or negative sign of the torque $\tau(x,y)$ defines whether the gas accelerates or decelerates.
An example of the gravitational potential and torque is presented in Fig.~\ref{fig_torque_pot} for NGC~6951.
The link between torque field and angular momentum variations is made through the observed distribution of gas \citep{Gar05} as explained in the following. At first we assume that the CO and HI emission lines are good tracers of the total gas column density in the central and outer disk, respectively. The reason for using two different lines is due to the fact that the neutral interstellar medium (ISM) undergoes a phase transition from atomic to molecular gas towards the center of a galaxy \citep{Young}. Since the gas distribution is the convolution of the gas density with the orbit path density, the gas distribution is indirectly equivalent to the time spent by the gas clouds along the orbit paths. Thus, we implicitly average over all possible orbits of gaseous particles.
To do this link between the derived torque field and angular momentum variations, the torques are weighted with the gas column density $N(x,y)$ and averaged over the azimuth: 
\begin{equation}
\tau(R) = \frac{\int_{\theta} \left[  N(x,y) \cdot (xF_y - yF_x) \right] } {\int_{\theta} N(x,y)},
\end{equation} 
By definition, $\tau(R)$ represents the azimuthal averaged time derivative of the specific angular momentum $L$ of the gas,
i.e.
\begin{equation}
\tau(R) = \frac{\mbox{d}L}{\mbox{d}t} \vert_{\theta}. 
\end{equation} 
The positive or negative sign of $\tau(R)$ defines whether the gas may gain or loose angular momentum, respectively.
The fueling efficiency can be estimated by deriving the average fraction of
the gas specific angular momentum transferred in one rotation
($T_{rot}$) by the stellar potential, as a function of radius, defined as:
\begin{equation}
\frac{\Delta L}{L} = \frac{\mbox{d}L}{\mbox{d}t}\vert_{\theta} \, \cdot \, \frac{1}{L} \vert_{\theta} \, \cdot \,  T_{rot} = \frac{\tau(R)}{L_{\theta}} \, \cdot \, T_{rot}
\end{equation} 
where the azimuthal averaged angular momentum $L_{\theta}$ is assumed to be well represented by its axisymmetric average, i.e., $L_{\theta} = R \cdot v_{rot}$. The inverse of $\Delta L/L$ determines how long it will take (in terms of orbital time periods) for the gravitational potential to transfer the equivalent of the total gas angular momentum. \par
The mass inflow/outflow rate of gas per unit length d$^2M(R)$/(d$R$d$t$) (in units of M$_{\odot}$~yr$^{-1}$~pc$^{-1}$) is calculated as follows:
\begin{equation}
\frac{\mbox{d}^2 M(R)}{\mbox{d}R \mbox{d}t} = \frac{\mbox{d}L}{\mbox{d}t}\vert_{\theta} \, \cdot \, \frac{1}{L} \vert_{\theta} \cdot 2\pi R \cdot N(x,y)\vert_{\theta}
\end{equation} 
where $N(x,y)\vert_{\theta}$ is the gas column density of the atomic (HI) and molecular gas using the conversion factor from CO to H$_2$ typical for galaxy nuclei of
$X=N(H_2)/I(CO)=2.2 \cdot 10^{20}$~cm$^{-2}$~K$^{-1}$~km$^{-1}$~s \citep{Sol91}. Then, the integrated inflow/outflow rates d$M(R)$/d$t$ are derived by integrating from $R=0$ out to a certain radius $R$,
\begin{equation}
\frac{\mbox{d} M(R)}{\mbox{d}t} = \sum_0^R \left[  \frac{\mbox{d}^2 M(R)}{\mbox{d}R \mbox{d}t} \cdot \Delta R \right] 
\end{equation}
with  $\Delta R$ as radial binning size (in units of~pc).

\subsection{Robustness of the method: Parameters and errors}
\label{subsec:torque_error}
\begin{figure}[ht]
\begin{center}
\includegraphics[scale=0.34]{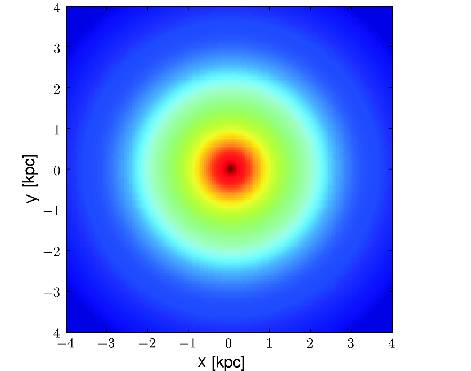}
\includegraphics[scale=0.34]{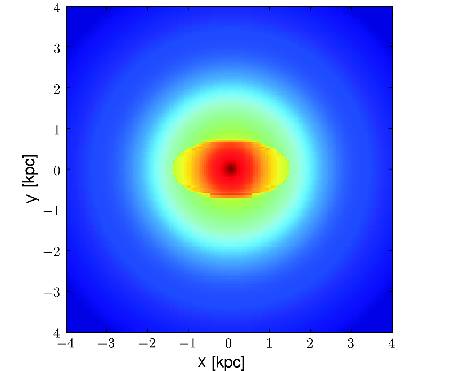}
\\

\includegraphics[scale=0.34]{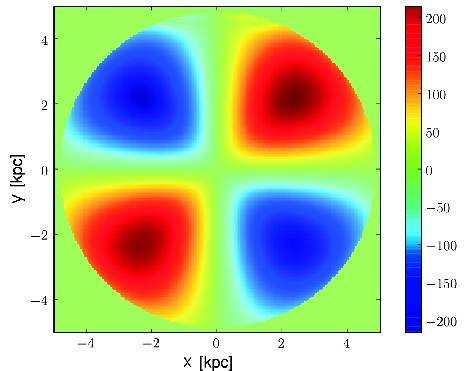}
\includegraphics[scale=0.34]{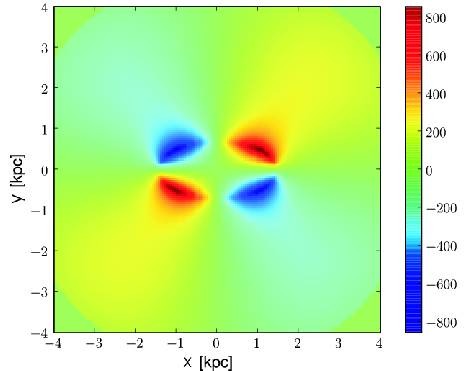}
\\

\includegraphics[scale=0.34]{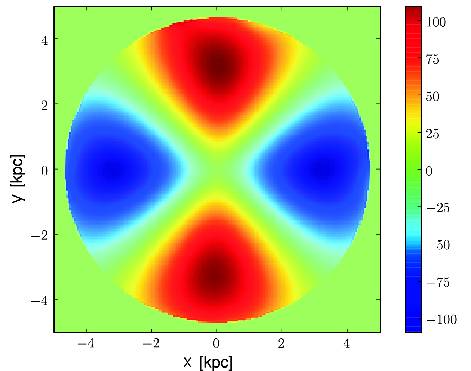}
\includegraphics[scale=0.34]{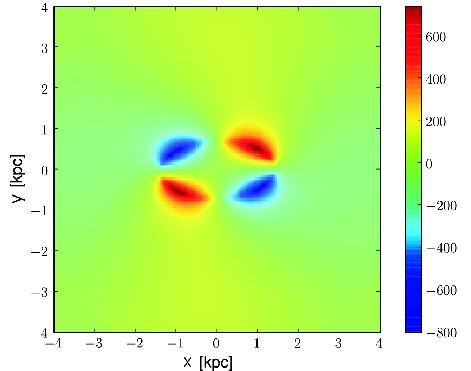}
\\

\includegraphics[scale=0.34]{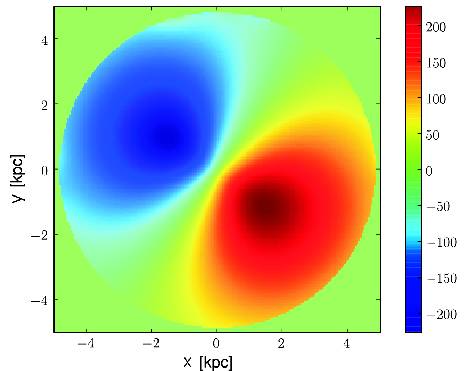}
\includegraphics[scale=0.34]{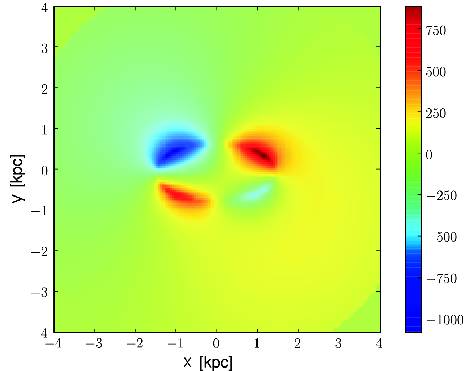}

\caption{\footnotesize Overview of the error estimation from models. The computed torques are calculated for two models: 1) an exponential disk (top left panel) which produces no torque pattern because of its axisymmetric distribution, and 2) an exponential disk plus an additional constant oval distribution (top right panel). The torque calculation is tested with uncertainties of an inclination error of 3$\degr$ (second row), a position angle error of 2$\degr$ (third row), and an error of the center position of 1 pixel $\simeq$ 1$\arcsec$ (bottom row).}
\label{fig_model}
\end{center}
\end{figure}

\begin{figure}[ht]
\begin{center}
\includegraphics[scale=0.4]{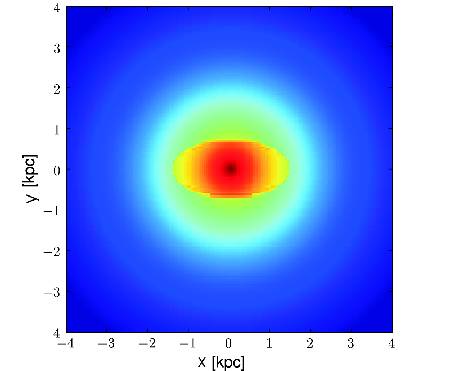}
\includegraphics[scale=0.4]{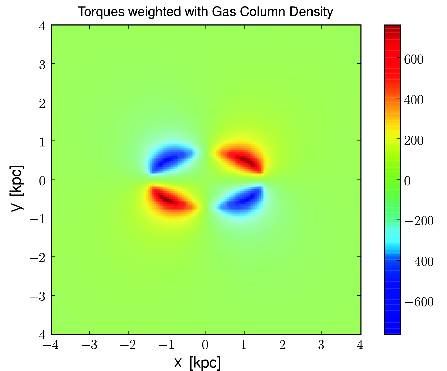}

\includegraphics[scale=0.4]{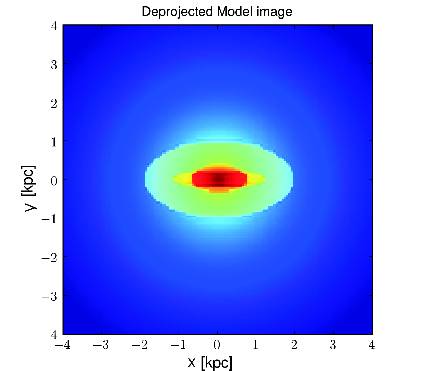}
\includegraphics[scale=0.4]{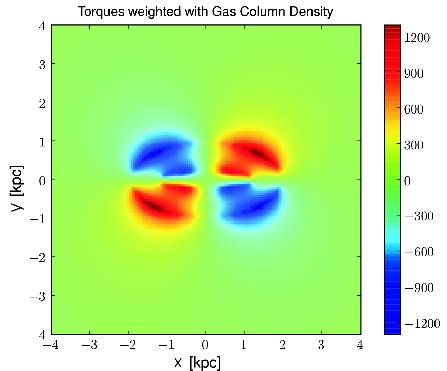}

\includegraphics[scale=0.4]{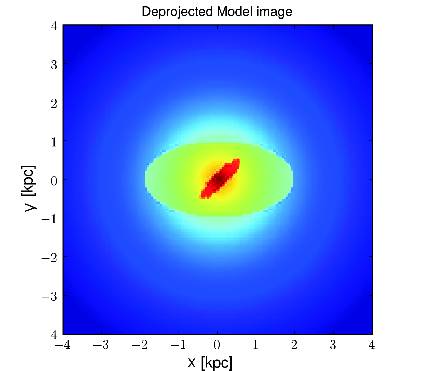}
\includegraphics[scale=0.4]{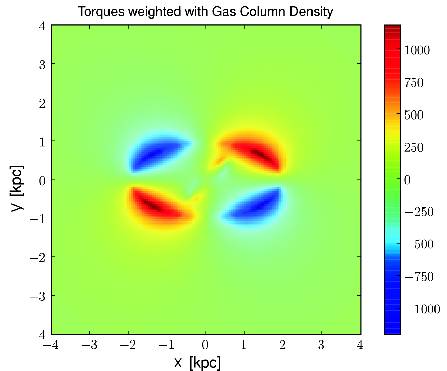}
\caption{\footnotesize Comparison of the torque patterns (right panels) derived from three different models (left panels): 1) exponential disk plus a constant oval distribution (top row), 2) exponential disk plus three parallel aligned constant oval distributions/bars with different lengths and strengths (middle row), and 3) exponential disk plus a constant oval distribution and an inner bar with a difference in PA of $45\degr$ (bottom row).}
\label{fig_3model}
\end{center}
\end{figure}

In order to examine the reliability of our results, we estimated how small errors in our assumed fitting parameters will affect the results. One main source for errors arises from uncertainties in the input parameters that were derived from our kinematic study (see previous section). 
First, we checked the parameters that describe the geometry of the disk given by inclination $i$, position angle PA, and position of the center $(x,y)$. We assume that these parameters do not vary with radius.  
To evaluate the effects of changes in these parameters on the derived gravity torque map, we tested two simple disk models (see Fig.~\ref{fig_model}): 1) an exponential disk which produces no torque pattern because of its axisymmetrical distribution, and 2) an exponential disk plus an additional constant oval distribution mimicking a barred galaxy, which produces a typical torque that changes sign between the 4 quadrants.  The oval distribution has a minor-to-major axis ratio of 0.5 and a constant surface brightness of 20\% of the maximum brightness of the exponential disk. For comparison, typical torque patterns without parameter errors are presented in Fig.~\ref{fig_3model} for different disk models. 
We tested the robustness of the results from our code with uncertainties based on our kinematic analysis (\S \ref{sec:kin}):
\begin{itemize}
\item An inclination error of 3$\degr$ at a typical inclination of 45$\degr$: The result shows an additional torque pattern that changes sign between the 4 quadrants. This pattern corresponds to one induced by a small oval contribution. Because of its axisymmetry and uniform gas distribution no contribution in the radial profile is seen. In principle an inclination error will produce a fake “bar” or “oval” but it will be along one of the principal axes of the disk (major, minor) and hence might provide a way to disentangle this false pattern from a true one.
\item A position angle error of 2$\degr$: Here the torque also changes sign between the 4 quadrants, causing a pattern similar to an oval distribution, but along an axis that is rotated by 45$\degr$ to one of the principal axes. 
\item An error of the center position of 1 pixel which corresponds to $\sim$50~pc at a distance of 10~Mpc and a pixel size of 1$\arcsec$: That results in a bimodal pattern, which increases the torque budget of one side of a torque pattern produced by an oval distribution, and decreases the torque at the opposite side.   
\end{itemize}
\par
\begin{figure}[ht]
\begin{center}
\includegraphics[scale=0.7]{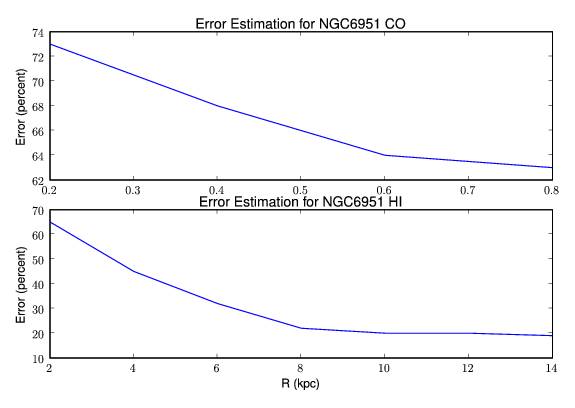}
\caption{
Error estimation for CO (top) and HI (bottom) based on radial torque calculation. The errors are the relative differences between the torque results using the best input parameters with those where typical errors of the input parameters were applied.}
\label{fig_test6951}
\end{center}
\end{figure}

Since this simple test does not take into account the gas distribution of a "real" galaxy which very likely differs from an homogeneous axisymmetric distribution, we also estimated the effects of these uncertainties using the NIR images and gas maps of a typical galaxy from our sample, namely NGC~6951.
In order to estimate the errors in the radial torques, we compared the torque results using the best input parameters with those where typical errors of the input parameter were applied. The total error as a function of radius is shown in Fig.\ref{fig_test6951}. The results of this test are summarized in Tab.~\ref{tab_errors}, separately for the central CO based and outer HI based torque calculation.
\par

Not only geometric uncertainties can have an effect on our results, but also the assumed mass-to-light ratio. Since we scale our potential with the derived (deprojected) rotation curve from the velocity maps, our results do not depend on uncertainties in the mass-to-light ratio, e.g. derived from stellar population models. Instead, our method relies on a scaling factor between the circular velocity derived from the stellar potential and the combined CO and HI rotation curve as given by Eq.~\ref{eq_scaling}.
Since rotation curves are derived from azimuthal averaged velocities, different mass-to-light contributions of non-axisymmetric components such as spiral arms (cool supergiants or dust glowing star formation regions) and bars will affect the scaling. Furthermore, possible deviations from a constant inclination of the galaxy disk (i.e. warping disk) might have an impact. However, most spiral galaxies show only small variations in the inclination across the stellar disk and we found no evidence for a significant warped disk in our sample as described in \S \ref{subsec:kin_obs}\par

Another possible source of error might be that the gas distribution is not 100\% recovered at all spatial scales due to missing short spacings of the interferometer observations. 
As our HI observations are continuous and traces all spatial scales, HI is a good tracer for the atomic gas distribution in our galaxies. For CO, the PdBI observations do not trace all spatial scales, and thus, might miss some flux from short spacings. However, to estimate the effect of short spacing corrections, \cite{Gar08} have compared the gravity torque results for NGC~4579 without (only CO PdBI-data) and with short spacing corrections using additional CO data obtained with the IRAM 30m telescope. The result shows that the gravitational torques with short spacings are roughly (5-40)\% larger than without short spacings.

\subsection{Determining the Corotation Radii}
\label{subsec:torque_phase}
The accurate determination of the corotation (CR) radii of density wave patterns is an important parameter for the characterization of the dynamical state of a galaxy and to understand the relation between morphologies and kinematics of galaxies. 
We have applied the \textit{Potential-Density Phase-Shift}  method \citep{Zha07} to derive the corotation resonance (CR) radii for our galaxies. This technique is based on the calculated radial distribution of an azimuthal phase shift between the stellar potential and a stellar density wave pattern which results in a torque action between the wave pattern and the underlying disk matter. Consequently, the material inside (outside) the CR radius looses (gains) angular momentum which causes an inflow (outflow).\par
The validity of the phase-shift method is based on the global self-consistency requirement of the
wave mode (i.e., both the Poisson equation and the equations of motion need to be satisfied at the same time).
For a self-sustained global spiral mode, the radial density variation of the modal perturbation density, as well as the pitch angle variation, together determine that the Poisson equation will lead to the zero
crossing of the phase-shift curve being exactly at the CR radius of the mode \citep[see for more details][]{Zha07}.
\par
We calculated the CR radius using the method of \cite{Zha07} as follows:
The rate of angular momentum exchange between the density wave pattern and the axisymmetric part of the disk can be either expressed as \citep{Zha96}:
\begin{equation}
\frac{dL}{dt}(r)=-\frac{1}{2\pi}\int_0^{2\pi}\Upsilon(r,\phi)\frac{\partial \Psi (r,\phi)}{\partial \phi}d \phi,
\end{equation} 
 with the perturbation density waveform $\Upsilon$ and the perturbation potential waveform $\Psi$, 
or as two sinusoidal waveform,
\begin{equation}
\frac{dL}{dt}(r)=\frac{m}{2}A_{\Upsilon}(r)A_{\Psi}(r)\sin\left[ m \phi_0(r)\right] ,
\end{equation}
with the amplitudes of the density wave $A_{\Upsilon}$ and potential wave $A_{\Psi}$, the non-axisymmetric mode number m (e.g. for two spiral arms or a bar: m=2), and the phaseshift $\phi_0$ between these two waveforms. Using these two equations, the phaseshift $\phi_0$ can be calculated from
\begin{equation}
 \phi_0(r,\phi)=\frac{1}{m}\sin^{-1}\left[ \frac{1}{m} \frac{\int_0^{2\pi}\Upsilon(r,\phi)\frac{\partial \Psi (r,\phi)}{\partial \phi}d \phi}{\sqrt{\int_0^{2\pi}\Psi^2 d \phi} \sqrt{\int_0^{2\pi}\Upsilon^2 d \phi}}  \right]. 
\end{equation} 
The potential $\Psi$ and the density distribution $\Upsilon$ are taken from our gravity torque calculation (see \S \ref{subsec:torque_desc}). We assume that the phaseshift is positive when the potential lags the density wave in the azimuthal direction in the sense of the galactic rotation. For all of our galaxies we derived the phaseshift as a function of radius and we defined the CR radii as the positive-to-negative crossings of the phaseshift $\phi_0$ (see Fig.~\ref{fig_phase}). At this location the direction of angular momentum transfer between the disk matter and the density wave changes sign.\par

In fact, the Phase-Shift method of \cite{Zha07} is mathematically very similar to our gravity torque study as both rely on a phase-shift between the gravitational potential and a density wave pattern. Thus, also the gravity torque is expected to change sign at each resonance with the gas as test particles for measuring the gravity torques. On the other hand, the gravity torque method might differ from the Potential-Density Phase-Shift in case of a superposition of density patterns or a shift of gas due to viscosity. However, in case of a strong dominating pattern, both methods are expected to reveal a similar location for the CR radius.\par

To verify our estimation of the CR radii we derived additionally the CR radii using the bar length and the method of \cite{Can93} (see \S \ref{subsec:kin_model}). A comparison of the results derived from these different studies is presented in Tab.~\ref{tab_bar} for all our galaxies. The CR radius is assumed to be 1.1-1.6 times the radius of the bar \citep{Rau08} and is visually estimated directly from the deprojected NIR images (Fig.~\ref{fig_torq_dpj}) using the change in PA and the ellipticity of the bar. All three methods (Canzian, bar length, Phase-Shift) reveal roughly the same radii for the CR of the bar. For our galaxies the Phase-Shift method appeared to be the most precise method with uncertainties of (5 - 10)\% and the Canzian method the most unprecise one with uncertainties ranging up to 50\%. The estimation of the CR using the bar length suffers from the uncertainty in the estimation of the bar length and its relation to the CR radius, which lies in the range of (10-30)\%. The comparison between the CR radii determined by our gravity torques study and the Phase-Shift method of \cite{Zha07} reveals significant differences for the location of the CR radii of the bar for NGC~3368, NGC4321, and NGC~6951 (see Tab.~\ref{tab_bar}). One possible explanation for these differences is that all three galaxies exhibit a strong stellar bar as well as a spiral pattern (also visible in the NIR images) which might be superimposed.  \par 

\begin{figure}[ht]
\includegraphics[scale=0.4]{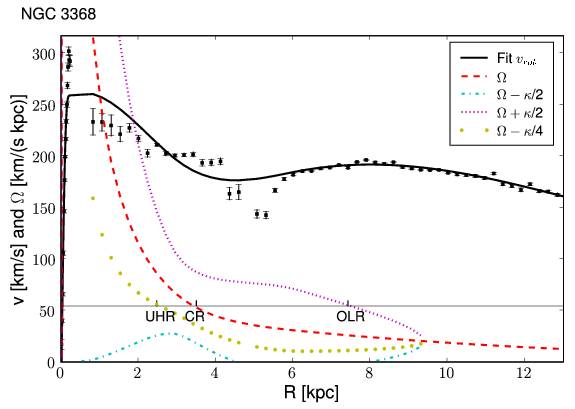}
\includegraphics[scale=0.4]{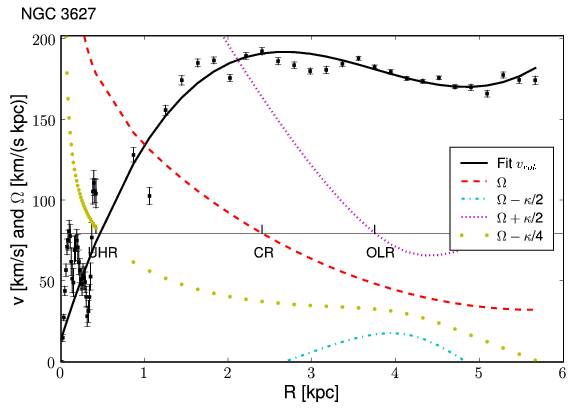}
\includegraphics[scale=0.4]{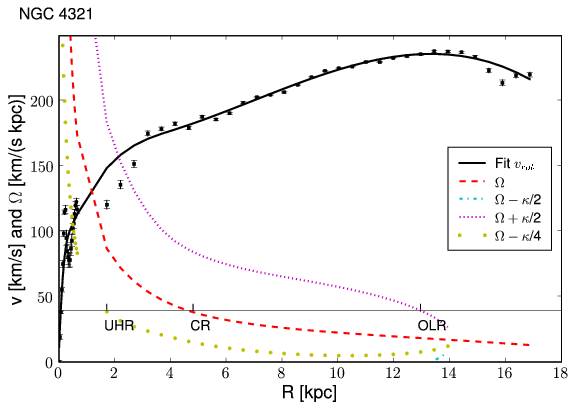}
\includegraphics[scale=0.4]{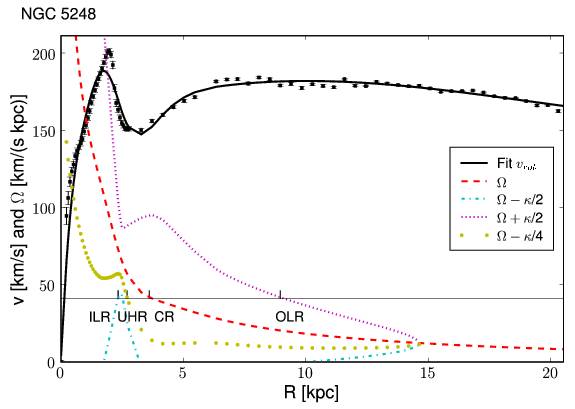}
\includegraphics[scale=0.4]{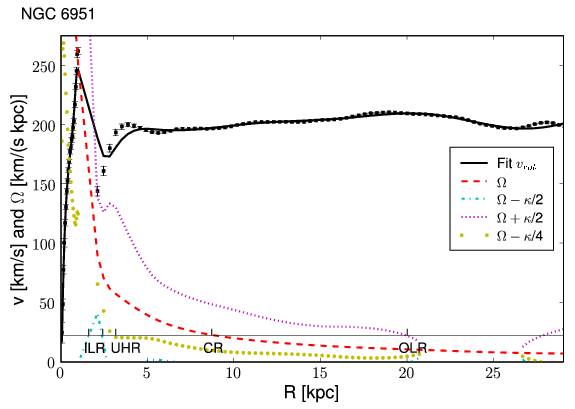}
\caption{ Overview of the CR resonance of the bar and the associated bar pattern speed  $\Omega_{Bar}$ (horizontal solid line, see also Tab.~\ref{tab_bar}) , the Inner Lindblad Resonance (ILR) at $\Omega_{Bar}=\Omega-\kappa/2$, the Ultra Harmonic-Resonance (UHR) at $\Omega_{Bar}=\Omega-\kappa/4$, and the Outer Lindblad Resonance (OLR) at $\Omega_{Bar}=\Omega+\kappa/2$ of the bar. The angular velocity $\Omega$ is calculated from a fit using a cubic spline interpolation (solid line) to our measured CO and HI rotation curve (points with errorbars). NGC~4736 and NGC~7217 are not shown here as no large-scale bar or oval is present in these two galaxies.}
\label{fig_resonances}
\end{figure}

To derive the location of the Inner Lindblad Resonance (ILR), the Ultra Harmonic Resonance (UHR), and Outer Lindblad Resonance (OLR) we used a simple method presented in Fig.~\ref{fig_resonances} which is described in the following. The angular velocity $\Omega$ is calculated from a fit using a cubic spline interpolation to our measured CO and HI rotation curve. After determining the bar pattern speed at the location of the corotation resonance (taken from the Phase-Shift method), the frequency curves $\Omega \pm \kappa/2$ and $\Omega - \kappa/4$ are derived with the epicyclic frequency 
\begin{equation}
\kappa = \sqrt{4 \Omega^2 + R \frac{\rm{d} \Omega^2}{\rm{d}R}}
\end{equation} 
The derived CR, ILR, UHR and OLR are also listed in Tab.~\ref{tab_bar}. Interestingly the OLR caused by the bar seems to overlap for some galaxies (NGC~3368, NGC~6951) with the CR radius of the spiral determined by the Phase-Shift method, suggesting a coupling between bar and spiral resonances. However, as our uncertainties in the estimation of the OLR are quite large, this correlation is not very significant.

\subsection{Results}
\label{subsec:torque_results}
We have derived neutral gas inflow rates for 7 nearby spiral galaxies as a function of radius as well as location within the disks (see top panel of Fig.~\ref{fig_torque}). By definition, the torque $\tau(R)$ represents the azimuthally averaged time derivative of the specific angular momentum of the gas. The positive or negative sign of $\tau(R)$ defines whether the gas may gain or loose angular momentum, respectively. For almost all galaxies of our sample the torque changes sign in different quadrants which corresponds to inflow and outflow in adjacent quadrants. As such a pattern is generally expected to be generated by m=2 modes, (e.g. bar or oval potential), we conclude that this mode is the dominant mode in our sample.
Only NGC~5248 shows a bipolar torque pattern for the center. In addition, the torque pattern is very sensitive to the presence of stellar spiral arms which cause a positive (negative) torque on the side of the spiral arm towards (away from) the galaxy center. Such a torque pattern is present in all galaxies of our sample, except for NGC~7217 and NGC~4736.  \par 

To estimate radial gas flows and time scales we derived the azimuthally averaged torque $\tau(R)$ and the angular momentum transferred in one rotation d$L(R)/L$ as a function of radius which are presented in Fig.~\ref{fig_torque}. Similarly to the torque maps, a positive or negative sign of $\tau(R)$ and d$L(R)/L$ defines whether the gas may gain or loose angular momentum corresponding to gas outflow or inflow, respectively. The transferred angular momentum and gas masses averaged over the entire galaxy disk are listed in Tab.~\ref{tab_torque} for each galaxy. We found that the average transferred angular momentum $\langle dL/L \rangle$ over the entire gas disk lies in the range of -0.04 (NGC~6951) to 0.06 (NGC~5248) for HI and -0.5 (NGC~3368) to 0.16 (NGC~3627) for CO. To estimate the mean amplitude of angular momentum exchange within the disk we derive the mean value of the absolute angular momentum transfer, $\langle \vert dL/L \vert \rangle$. This parameter is used later in \S \ref{sec:dis} for a comparison to the non-circular motions identified in our analysis of the gas kinematics. We found values of  $\langle \vert dL/L \vert \rangle$ in the range of 0.03 (NGC~7217) to 0.26 (NGC~3627) for HI and 0.03 (NGC~7217) to 0.55 (NGC~3368) for CO. Clearly, NGC~7217 exhibits the lowest torque strength and angular momentum transfer in our sample due to its nearly axisymmetric potential \citep{Com04} whereas NGC~3627 and NGC~3368 exhibit the largest amplitude of angular momentum exchange.
In general, the amplitude of $\langle \vert dL/L \vert \rangle$ is larger in the center (CO) than in the outer disk (HI). \par

To estimate the transferred gas mass at a certain radius, we derived the mass inflow/outflow rates d$^2M(R)$/(d$R$d$t$) (in units of M$_{\odot}$~yr$^{-1}$~pc$^{-1}$) as a function of radius (see Fig.~\ref{fig_torque_mass}). In addition, we examined the net mass flow d$M(R)$/(d$t$) (in units of M$_{\odot}$~yr$^{-1}$) outwards to a given radius $R$ by integrating over the gravitational torques of the molecular and atomic gas disk, separately. These values represent the net gas mass inflow/outflow within these regions and can be compared to the typical gas masses required by star formation and AGN fueling (see \S \ref{sec:dis}). A positive (negative) sign indicates gas outflow (inflow).
The net gas mass flow for the entire molecular and atomic gas disks (integrated from 0 to $R_{max}$) are listed in Tab.~\ref{tab_torque} with a range of (-0.01 - 0.5)~M$_{\odot}$~yr$^{-1}$ for the atomic gas ($R_{max}\simeq 20$~kpc) and much larger values of (-23 - 50) M$_{\odot}$~yr$^{-1}$ for the molecular gas in the center ($R_{max}\simeq 1$~kpc). 
However, to study the redistribution of gas within the galaxy disk, e.g. caused by bar and spiral wave patterns, one has to examine the gas mass flow as a function of radius (see Fig.~\ref{fig_torque_mass}). For example, the net gas mass flow over the entire disk can be zero while a large redistribution of gas may occur within the disk. This seems to be the case for, e.g., NGC~3627 where we found a large net gas mass inflow in a region of $0<R<3$~kpc ($\sum_0^{3~kpc}$~M$_{\odot}$~yr$^{-1}$=-1~M$_{\odot}$~yr$^{-1}$), and a similar value with the opposite sign (gas mass outflow) from 3~kpc to the outer region of the galaxy ($\sum_{3~kpc}^{6~kpc}$~M$_{\odot}$~yr$^{-1}$=1~M$_{\odot}$~yr$^{-1}$), so that the sum over the entire disk results coincidentally in an almost zero net gas mass flow ($\sum_0^{R_{max}}$~M$_{\odot}$~yr$^{-1}$=-0.1 M$_{\odot}$~yr$^{-1}$).
In general, we found for our sample that the amplitude of the gas mass flow within the disk is much larger than the net gas mass flow over the entire gas disk. This indicates that the redistribution of gas within the galaxy exceeds the net gas mass outflow/inflow from the entire galaxy. The significant inflows and outflows within the disk are likely caused by the dynamical action of a stellar bar and/or spiral pattern. \par

To evaluate in more detail whether the angular momentum transfer changes sign at characteristic dynamical locations within the disk, we derived the corotation (CR) radii as described in \S \ref{subsec:torque_phase} (see Tab.~\ref{tab_torque} for an overview of all CR radii for our galaxies). The CR radius is defined as the radius where the density wave pattern (e.g. bar, spiral) and the differentially rotating disk have the same angular velocity. This leads to a gain of angular momentum outside and loss of angular momentum inside the regions close to the CR radius. Since galaxies can harbor several spiral and/or bar patterns, also multiple CR radii can exist within one galaxy. In Tab.~\ref{tab_torque} we listed all CR radii for our galaxies with a significant phaseshift inversion. \par 
 
Summarizing, almost all galaxies of our sample (except the center of NGC~5248) exhibit torque patterns typical for m=2 modes (bars, spirals). NGC~3627 seems to have a very strong torque efficiency in the center (d$L/L>0.6$) while NGC~7217 shows the lowest torque strength and angular momentum transfer in our sample. In general, we found that the transfer of angular momentum and the gas mass flow is larger in the center ($R\lesssim 1$~kpc) than in the outer disk ($R \gtrsim 1$~kpc).

\section{Discussion}
\label{sec:dis}

\subsection{Gas flows in individual galaxies}
\label{subsec:dis_galaxies}
In this section we describe the gas flow for each of our 7 galaxies and search for relations to their dynamical states and stellar and gaseous morphologies. As most of our galaxies are different in their morphology and kinematics, only case studies can provide information about the mechanisms that are acting. A general picture of the gas flow follows from a comparison of these individual studies and is described in \S \ref{subsec:dis_scale}.

\subsubsection{NGC~3368}
NGC~3368 has two stellar bars, an inner one with a radius of $\sim$1.0~kpc and an outer one with a radius of $\sim$3.2~kpc, which have a small offset of $\sim 25 \degr$ between their position angles (see Fig.~\ref{fig_torq_dpj}). Two spiral arms are connected to the ends of the outer bar. The atomic gas distribution shows two gaseous spiral arms which are coincident with the stellar ones and an additional gaseous ring around the outer bar. We found that the two dimensional torque maps of the central and the outer disk show the typical torque pattern that is expected from their stellar bars and spirals. On the basis of our radial torque profiles we found gas inflow for the outer disk region (7-10~kpc) with an average angular momentum transfer d$L/L$ of 10\% of the total angular momentum per rotational period at a given radius, while on intermediate disk scales (2-7~kpc) gas outflow is present (see Fig.~\ref{fig_torque}). Because of the switch from out- to inflow, gas is expected to accumulate at this radius which is confirmed by our observed atomic gas distribution (see Fig.~\ref{fig_torque_mass}). Interestingly, the shift at 7~kpc in the flow direction is directly contrary to the one that is expected by crossing the CR resonance of the spiral at $\sim$7~kpc. The lack of a correlation between CR radius of the spiral and the expected gas flow can be likely explained by the fact that the bar potential is still acting at this location and overlaps with the spiral pattern. The UHR ($R\simeq 2.5$~kpc) and OLR ($R\simeq 7.5$~kpc) of the bar seem to correlate with an expected accumulation of gas at $R= 2.5$~kpc and $R= 8$~kpc, respectively, also clearly visible in the radial atomic gas density profile (see Fig.~\ref{fig_torque_mass}). For the center we found a large gas inflow from 2 - 0.08~kpc, that seems to transport the gas towards the AGN. Only for the very center from 0 - 0.08~kpc gas outflow is present, suggesting an accumulation of gas at a radial distance of 0.08~kpc, which is also visible as a small bump in the radial molecular gas density profile (see Fig.~\ref{fig_torque_mass}). 

\subsubsection{NGC~3627}
The stellar distribution of NGC~3627 exhibits two bars with a radius of 1.5~kpc and 0.2~kpc and two spiral arms at the ends of each of the bars. As the two bars have an offset between their PAs of $\sim45 \degr$  a possible decoupling of these two dynamical modes might influence the gas kinematics \citep[see also][]{Gar08}. We found that the torque pattern reflects the shift of the PAs between these two bars. For the outer disk the torques are dominated by the potential of the spiral arms with a CR radius at 2.5~kpc. Within this CR radius the gas is transported inwards, while moving outwards outside the CR radius, which is confirmed by our radial torque profile: A significant inflow (d$L/L\simeq0.3$) from 2.8-0.7~kpc and outflow of gas (d$L/L\simeq0.3$) from 2.8-5~kpc. For the central disk traced by our CO observation we found a large gas outflow (d$L/L\simeq0.4$) from 0 - 0.5~kpc. Thus, gas accumulation might be possible at $\sim0.5$~kpc where the flow direction switches from out- to inflow, which is also visible as a small peak in the radial molecular gas density profile (see Fig.~\ref{fig_torque_mass}). This gas accumulation at $\sim 0.5$~kpc seems also to overlap with the UHR of the outer bar at $R= 0.45$~kpc.     

\subsubsection{NGC~4321}
The inner region of NGC~4321 has been extensively studied in the NIR \citep{Kna95a, Kna95b} and shows a prominent nuclear bar aligned with the large-scale stellar bar. The large-scale bar is coupled to 4 spiral arms. The gas flow follows the spirals in the outer disk and is increasingly dominated by the torques caused by the bar towards the center. For the outer disk (from 12 - 18~kpc) we found a significant outflow (d$L/L=0.2$) and on intermediate scales (from 12 - 6~kpc) inflow. The switch from inflow to outflow is roughly occurring at the CR radius of the spirals at 11.2~kpc. At the CR radius of the bar no switch from in- to outflow is found, suggesting that the standard CR determinations are not applicable for this case because of a superposition of bar and spiral wave pattern. While a shift from out- (0.5 - 0.9~kpc) to inflow (1.0 - 0.9~kpc) is present at 0.9~kpc, no significant angular momentum transport occurs towards the very center of the galaxy (from 0.5 - 0~kpc). The comparison with a previous gravity torque study only for the center \citep{Gar05} revealed a similar torque pattern but larger amplitudes. Although similar methods have been used, the difference in the amplitude can be likely explained by differences in the geometric parameters used for the disk ($i$, PA, and center) as well as a different method for the scaling with the mass-to-light ratios (\cite{Gar05}: constant scaling; our study: scaling as function of radius). 

\subsubsection{NGC~4736}
For NGC~4736 we found that the direction of the torque rotates by about $\sim60 \degr$ in the region of 2 - 0~kpc. This is likely due to the presence of two oval stellar distributions whose major axes have an offset of $\sim 60\degr$ suggesting a decoupling between these two dynamical modes. These ovals are clearly visible in the NIR images (see Fig.~\ref{fig_torq_dpj}). The gravitational torque caused by the outer oval is still dominant in the region of the large gas spiral at 6.2~kpc producing a net outflow at this location (5.5 - 7~kpc). Averaged over the entire atomic gas disk, gas is transported outwards.  

\subsubsection{NGC~5248}
NGC~5248 shows a prominent stellar large-scale bar \citep{Jog02a} visible in the NIR image and in the gaseous distribution as well \citep{Haa08}. Further, two inner stellar spiral arms are present within the bar \citep{Jog02} and two outer ones continue from the ends of the bar to the outer disk at 15~kpc. The gravitational torque pattern seems to be a mix between the bar and spiral arm contributions.  
In particular, the outer disk exhibits an additional torque concentration at the end of the north-western spiral arm at a distance of $~7$~kpc from the center. 
For the outer disk we found outflow at distances from (11-15), (2-8)~kpc and inflow between $\sim$(11-8)~kpc.
The gravitational torques in the center are very small (maximum of d$L/L=0.04$) and show no typical torque pattern for a bar or oval. Instead a bipolar pattern is present, but might be caused by a possible offset between the real center and our used values (see \S \ref{subsec:torque_error}) or is alternatively suggesting the presence of a lopsided disk.  
Interestingly, NGC~5248 is the only galaxy in our sample without AGN activity, which might be linked to the absence of gravitational torques and/or a possible lopsided disk in the center. However, to test this finding in detail higher spatially resolved data of the molecular gas distribution with $\sim0.5\arcsec$ is necessary as for this galaxy only BIMA-SONG data with $6.1\arcsec$ was available for our study. 

\subsubsection{NGC~6951}    
NGC~6951 is a representative galaxy with a combination of AGN activity, stellar bars (inner and large-scale) and spiral arms (visible in NIR image and HI map). The gas flow pattern from 15-28~kpc reflects the stellar spiral arms and is increasingly dominated by the gravitational torques caused by the bar towards the center (see Fig.~\ref{fig_torque}). The molecular gas distribution in the center shows two gaseous spiral arms whose gas flow underlie the gravitational torques caused by the inner bar. For the outer disk we found that the gas is transported outwards (d$L/L\simeq0.15$) from (19-27)~kpc. On intermediate scales inflow occurs from (19-11)~kpc, while outflow is dominant from (4-11)~kpc. The radius where the gas flow is changing from inflow to outflow (11~kpc) does not overlap with any resonances of a single pattern (i.e. bar or spiral). We interprete this as gas is funneled inwards across the CR radius of the bar (at $\sim$8.5~kpc) and that the gas can overcome the CR barrier. We can exclude this as being due to errors in the gravitational torque calculation or CR estimation, as neither the errors of our torque study (see \S~\ref{subsec:torque_error}) are large enough to eliminate such a significant inflow on the scale between (5-11)~kpc, nor the possible range in the CR estimation of (5-9)~kpc (including errors) could shift the CR radius to the switch from in- to outflow at 11~kpc. Also viscous torques are not efficient enough at these large scales to change the flow direction, because of much smaller gas densities than in the center. The most likely explanation for the mismatch between the gravitational torque induced flow direction and the standard CR estimations of the bar is the presence of several pattern speeds (i.e. for bar and spiral wave pattern) that combine and allow the gas to overcome the standard CR barrier of a single pattern.
For the center we found gas inflow from (0.8-0.45)~kpc and outflow from (0.3-0.45)~kpc, suggesting a gas accumulation at $\sim$0.45~kpc. This is roughly coincident with a significant peak in the radial gas density profile at this location (at 0.38~kpc) leading to two inner molecular gaseous spiral arms. The comparison with a previous gravity torque study only for the center \citep{Gar05} revealed a similar torque pattern but much larger amplitudes than our study (about a factor of 5-10). This difference in amplitude might not only be caused by differences in the used geometric parameters of the disk ($i$, PA, and center) as well as a different method for the scaling with the mass-to-light ratios, but eventually also by the use of different NIR images (\cite{Gar05}: J-band; our study: K-band). We tested our code with the NIR image and parameters used by \cite{Gar05} and found similar torque values as in \cite{Gar05}, so that an error in our code can be excluded.   

\subsubsection{NGC~7217}
The two-dimensional torque map shows a pattern typical for an oval potential. Although NGC~7217 has no bar, an oval stellar distribution is possible, however it might be just an deprojection effect due to uncertainties in the inclination angle. The latter case seems to be plausible as the torque pattern is aligned with the major axis (PA=265.5$\degr$) in such a way as expected from an error in inclination angle (see \S \ref{subsec:torque_error}). For the outer disk we found no strong evidence for gas transport caused by the gravitational torque. At the center we found only small evidence for gas inflow from (0.5-0.2)~kpc with a maximum of d$L/L=0.1$, but this is presumably due to the uncertainty in the inclination angle.

\subsection{A general picture of the gas flow from the outer disk to the center}
\label{subsec:dis_scale}

We found that the typical torque pattern changes sign in 4 quadrants. This typical butterfly pattern is characteristic for the action of an oval or barred gravitational potential. The orientation of the torque pattern in the center is coincident with the one of the outer disk (except for NGC~4736). Towards the outer disk, the gas flow is governed by the action of the stellar spiral arms, if present, which cause a positive (negative) torque on the side of the spiral arms towards (away from) the galaxy center. Thus, we conclude that for almost all galaxies in our sample a dominant m=2 mode is present and has roughly the same orientation in the outer disk and the center. We found nested bars within large-scale bars for NGC~3368, NGC~3627, NGC~4736 whose major axes have an offset of $25\degr - 60\degr$ to the large-scale bars.
No signs for unidentified patterns (i.e. not showing up as a prominent feature in the NIR images) are found in the torque maps.
\par
The radial profiles of d$L$/L and d$^2M(R)$/(d$R$d$t$) reveal that torques are more efficient in the redistribution of gas in the center of galaxies than in the outer disks, presumably due to larger gravitational torques induced there by the bar component. We found for a majority of our galaxies (5/7) a significant reversal from outflow to inflow in radial direction at radial distances ranging between (0.1 - 0.9)~kpc, leading to an increasing gravitational pressure on the gas at this position due to gravitational torques and in case of stable orbits to an accumulation of gas at this position. A comparison of our radial gas density profiles and the torque profiles revealed for most of the outer disks (HI density) no correlation while for some central disks (CO density) an overlap or a shift between both profiles is found (see Fig.~\ref{fig_torque_mass} and \S \ref{subsec:dis_galaxies}). A possible explanation is that gravitational torques will redistribute the gas within a few orbital rotation periods. As the transport of angular momentum per rotation period is on average $\sim$0.3, the expected timescale for the redistribution of the total angular momentum for a given radius is $\sim$3 rotation periods  ($\cong 3\cdot 10^7$~yr at a radius of 0.5~kpc). Another explanation might be that the gas is consumed in star formation and hence does not accumulate as expected. 
For the very center (0 - 0.1)~kpc we found no significant gas transport towards the AGN due to gravitational torques, suggesting that other torques such as viscous torques might be become important for fueling the AGN \citep[see also][]{Gar05}.  
\par 
In the dynamics of galaxies the CR radius is expected to separate inflow (between ILR and CR) from outflow (between CR and OLR). However, in all galaxies of our sample the CR radius of the bar does not correlate with any inflow/outflow separation and thus, sets no barrier for gas transport within the disk. This contradiction can be possibly explained by a coupling of dynamical modes, e.g. of a spiral and a bar or an oval potential. 
Thus, the orbital paths of the gas may change in comparison to the orbits of a single dynamical mode and allow the transport of gas across the CR. A similar case has been evolved for NGC~4579 which shows a transportation of gas inwards across the CR of the bar, but in this case, due to (morphological) decoupling of bar and spiral pattern on different spatial scales \citep{Gar08}.\par

With regard to the CR of a spiral pattern, most of our galaxies with spirals (NGC~3627, NGC~4321, NGC~5248) show a correlation between CR radius of the spiral and a reversal of inflow/outflow of gas due to gravitational torques. Only for NGC~6951 no correlation is found. Instead a significant outflow due to gravitational torques occurs at the CR radius of the spiral, suggesting that the bar potential is still acting at this radius and may overlap with the one of the spiral pattern. Further, the OLR of the bar (18.6~kpc) does not overlap with the CR of the spiral and, thus, seems to substantiate the dominance of the bar for NGC~6951. Also for NGC~3368, NGC~4736, and NGC~7217, which have none or only very weak stellar spiral arms, the gravitational torque is dominated by a bar or oval potential.
\par   
Summarizing, gravitational torques can be very efficient in transporting the gas from the outer disk to the center at$\sim$100~pc. Due to a dynamical coupling of bar and spiral arms, the gas can bridge even resonances such as the CR of a bar (which could act as a natural barrier for further gas inflow).

\subsection{Comparison between gravitational and kinematic derived gas flows}
\label{subsec:dis_comp}

\subsubsection{Estimation of radial motion}
In general one would expect that gas inflow/outflow should also be recognizable in the velocity field of a galaxy. 
In practice, to determine radial gas flow from the line-of-sight velocity alone is very ambitious and controversial \citep[for an overview see][]{Won04}.
The main reason is that the dominant contributor to noncircular motions (see next section) in a bar or spiral potential seems to be elliptical streaming which cannot be used by itself to derive net inflow/outflow without further assumptions. One of the most promising attempts to derive radial gas motion was conducted by \cite{Won04} using the variations of the first- and third-order sine coefficients ($s_1$ and $s_3$) as a function of radius to provide a basic approach to diagnose various types of non-circular motions \citep[see for more details][]{Won04}: warp streaming (i.e. motion out of the plane of the disk; correlation between $s_1$- and $s_3$-term, with a slope d$s_3$/d$s_1>0$), elliptical or spiral streaming (anticorrelation between $s_1$- and $s_3$-term, with d$s_3$/d$s_1<0$), and radial inflow ($s_1 \gg s_3$). Note that these criteria are very idealized and do not permit a clear distinction between the models when applied to real galaxies. 
However, to definitely test whether this method can clearly identify radial gas motion or not, we have compared the radial gas flows derived from our velocity fields (see \S \ref{subsec:kin_model}) with the inflow/outflow rates due to gravity torques (see \S \ref{subsec:torque_results}). \par

We have used the following convention \citep[following the method of][]{Won04}: If the $s_1$-term is interpreted as radial gas flow (for $s_1/s_3\gg 1$), a positive (negative) sign of the $s_1$ amplitude corresponds to gas outflow (inflow) for a counterclockwise rotating galaxy. We derived the sense of rotation for our galaxies by presuming that spiral arms seen in the optical images are always trailing. In order to have a consistent labeling we defined $s_1^*= s_1$ ($ -s_1 $) for counterclockwise (clockwise) rotation, so that $s_1^*<0$ corresponds to gas inflow (see third panel of Fig.~\ref{fig_harmonics}). 
In the following we discuss some candidate regions for inflow/outflow:  NGC~3627 and NGC~4321 exhibit a significant negative $s_1$ term at 0-0.15~kpc and 0.2-0.8~kpc, respectively, which indicates radial inflow at these scales. In  contrast, we found for these regions significant outflow rates from our gravitational torque analysis. Further, the outer disk of NGC~6951 shows a significant negative $s_1$ term (indicates inflow) at (15-18)~kpc, while the torque analysis reveals significant outflow (d$L/L=0.1$). Summarizing, we found no correlation between the candidate radial flow regions from our kinematic study and the inflow/outflow caused by gravitational torques. Thus, we conclude in agreement with other studies \citep[e.g.][]{Won04} that the non-circular terms are predominantly caused by elliptical streaming motions and cannot be used as indicator alone for possible radial inflow/outflow of gas.

\subsubsection{Streaming motions versus radial inflow/outflow}
As an alternative to radial gas motion where the gas looses or gains angular momentum, gas clouds with streaming motion can follow under certain assumptions closed orbits and no angular momentum is transferred when collisional shocks and other nonlinear effects are excluded.   
For instance, if the principal axes of an elliptical orbit are aligned with those of a stellar bar, the 
phase shift between the bar and the gas is zero and hence no inflow will occur at all, but still strong 
streaming motions are present. To produce a net inflow/outflow a phase shift between the gas and the potential is required. 
Streaming motions are present in m=2 perturbation such as in a barlike potential where the gas follows elliptical closed orbits or in a two arm spiral density wave where a phase shift occurs that varies as a function of radius. This collisionless orbit approximation can be simple described by first order Fourier terms \citep{Fra94}.
In a realistic treatment the nonlinear effects will contribute additional Fourier terms and elliptical streaming motions may overlap with radial gas flow.\par 

To estimate a possible dependence of streaming motion on pure inflow/outflow rates, we have conducted the following study. If we assume that the non-circular motions derived from our kinematic harmonic decomposition are dominated by elliptical streaming, it is possible to compare the amplitude of non-circular motions $v_{nc}/v_{rot}$ as tracer for elliptical streaming with the amplitude of angular momentum transport caused by gravitational torques $\langle \vert \mbox{d}L/L \vert \rangle$ as a function of radius (see Fig.~\ref{fig_comp}). Note that the absolute scales are not really expected to be similar as they quantify different physical measurements, but indicate the significance of the values obtained with each method.\par

On average, we find that the amplitude of angular momentum exchange derived from our torque analysis ($v_{nc}/v_{rot}=0.09$ for HI and $v_{nc}/v_{rot}=0.16$ for CO) is similar to the amplitude of non-circular motions obtained from our kinematic analysis ($v_{nc}/v_{rot}=0.09$ for HI and $v_{nc}/v_{rot}=0.26$ for CO). In more detail, $v_{nc}/v_{rot}$ seems to be larger for the center as $\langle \vert \mbox{d}L/L \vert \rangle$ for most of our galaxies, except for NGC~3368 and NGC~6951. For the outer disk $\langle \vert \mbox{d}L/L \vert \rangle$ is larger than $v_{nc}/v_{rot}$, at least for those galaxies with a bar. Galaxies without a bar (NGC~4736 and NGC~7217) exhibit very weak gravitational torques and hence small values of $\langle \vert \mbox{d}L/L \vert \rangle$. Interestingly, non-circular motions in unbarred galaxies have roughly the same amplitude for the outer disk than barred galaxies. \par

The peaks in the amplitude of non-circular motions to the amplitude of inflow/outflow seem to be correlated as a function of radius for some galaxies: Center and outer disk of NGC~3368, outer disks of NGC~3627, NGC~4321, NGC~4736, and NGC~6951. Hence one might infer that non-circular motions are not pure elliptical streaming motions, but include to some extent radial motion as well. On the other hand, we found also cases with significant streaming motions but negligible inflow/outflow rates (e.g. for NGC~4736 and NGC~7217). Thus, our study demonstrates that signatures for streaming motions alone can not be used as reliable tracers for inflow/outflow which requires a phaseshift between the gas and the gravitational potential.

\subsection{Implications for AGN fueling}
\label{subsec:dis_AGN}
The fueling of AGN requires transportation of gas from large~kpc scales down to the inner~pc. To bring the gas to the center the angular momentum of the gas must decrease by orders of magnitude.
Several types of gravitational instabilities, such as bars, nested bars, and spiral density waves have been proposed that might combine to transport gas from large~kpc scales down to the very center \citep[e.g.][]{Shl90, Com03, Wad04}. Inflow of gas is expected to occur mainly inside the corotation radius (CR), transporting the gas towards
the inner Lindblad resonance (ILR), where the gas might accumulate and start to form clumps of star formation.
To test this scenario we have compared the gas inflow/outflow rates to the dynamical behavior of the gas at the resonances. Our results suggest for some galaxies of our sample (NGC~3368, NGC~4321, and NGC~6951) that the CR radius of the bar is overcome by gravitational torques, most likely due to a coupling of several pattern speeds (e.g. spiral arms plus bar wave pattern). Thus, the gas reservoir inside the CR can be refueled by gas from the outer disk. 
However, a generalization of such a scenario is not possible and detailed case studies are required to disentangle the underlying mechanisms. For example, \cite{Gar08} find for NGC~4579 that a decoupling of bar and spiral patterns on different spatial scales allows the gas to efficiently populate the UHR region inside corotation of a bar and thus explain how the CR barrier can be crossed. \par

For the central disk, we found that the integrated gravitational torques indicate net outflow (2 galaxies) and inflow (5 galaxies) in our galaxies. In particular NGC~3627 (Seyfert 2 type) shows a large gas outflow (d$L/L$=0.5) inside the inner 0.4~kpc, suggesting a redistribution of the total angular momentum of the gas in $\sim$2.5 rotational periods at this radius ($\backsimeq 2.4 \cdot 10^7$~yrs). 
Interestingly, the kinematic analysis reveals also a large amplitude of non-circular motions (d$v_{nc}/v_{rot}$=0.8) at scales of $R\leq0.15$~kpc which may substantiate this scenario. \par

These findings are similar to other studies of the gas flow for the molecular gas disk: A gravity torque study for NGC~6574 using CO as tracer \citep{Lin08} has shown that gas is flowing inwards down to a radius of 400~pc which overlaps with a high nuclear gas concentration suggesting that the gas has been piling up at this location quite recently, since no starburst has been observed yet. An analysis of the torques exerted by the stellar gravitational potential on the molecular gas in four other (NUGA) galaxies found a mostly positive torque inside r$<$200 pc and no inflow on these scales suggesting that viscous torques may act as fueling mechanism as well \citep{Gar05}.
An explanation of this variety of morphologies might be related to timescales \citep{Gar05}. In this scenario the activity in galaxies is related to that of bar instabilities, expecting that the active phases are not necessarily coincident with the phase where the bar has its maximum strength. The resulting implication is that most AGN probably occur between active accretion episodes of the galaxy disk. Evidence for such a scenario was found by \cite{Hun08} for NGC~2782 where molecular gas inside the ILR of the primary bar, transported by a second nuclear bar, suggests that the gas is fueling the central starburst, and in a second step might directly fuel the AGN.\par

The integrated gas mass transfer rates in the center range from 0.01 - 50~M$_{\odot}$~yr$^{-1}$ based on our assumptions (see \S \ref{subsec:torque_desc}) and are sufficient enough to preserve AGN activity which requires typical mass accretion rates of 0.01-0.1~M$_{\odot}$~yr$^{-1}$ for LINER and Seyfert activity. However, to bridge the last 100~pc to the AGN other mechanisms than gravity (e.g. viscous torques) have to play a major role, since no significant gravity torques are present at these scales \citep[see also][]{Gar05}.

\section{Summary}
\label{sec:sum}
We studied the gas flow in 7 nearby galaxies to quantify the gas flow towards the center in AGN galaxies.
On the basis of high angular resolution CO and HI spectroscopic data and NIR images we calculated the gravitational torques and inflow rates as a function of radius and location within the galaxies. 
Besides the two-dimensional gravitational torques, we applied for this study a variety of methods including a harmonic decomposition of the residual velocity field, and the Canzian and Phase-Shift methods to identify dynamical resonances. Our study shows that gas redistribution is very effective within the galaxy disks due to gravitational torques indicating that spiral disks are very dynamic systems that undergo strong radial evolution on timescales of a few rotation periods (e.g. $\sim 5 \cdot 10^8$~yrs at a radius of 5~kpc).
The main results are summarized as follows:
\begin{itemize}
\item Gravitational torques are very efficient in transporting the gas from the outer disk to the center ($\sim$100~pc). Typical rates for cold gas angular momentum transport per rotation are 10\% of the total angular momentum at a given radius. The transported gas mass rates range from 0.01 to 50~M$_{\odot}$~yr$^{-1}$. 
\item Dynamical resonances such as the Corotation Resonance (CR) of a bar pattern, which could act as a barrier for gas transport, are apparently easily overcome by the gas flows that are induced by gravitational torques. A possible explanation is the presence of several patterns with different speeds (e.g. bar and spiral wave pattern) that combine and allow the gas to overcome the standard CR barrier of a single pattern.
\item The transport of angular momentum, and thus, the inflow/outflow rates, are larger for the central ($<1$~kpc) than for the outer disk (1-20~kpc), suggesting stronger gravitational torques induced by bars in the center of galaxies.
\item In our sample the dominant dynamical mode is m=2 (e.g.bar,oval,two-armed spiral-arms) and has roughly the same orientation across the entire radial range for the outer disk than for the center (except for NGC~4736).
\item Our gas flow maps indicate the presence of nested bars within larger bars for 3 out of 7 galaxies (NGC~3368, NGC~3627, and NGC~4736) with offsets in the position angles of $25\degr - 60\degr$ to the major axes of the large-scale bars.
\item No kinematic signs for unidentified non-axisymmetric mass distributions (that are not evident in the NIR images) are found in the torque maps.
\item Non-circular motions seem to correlate only in a few cases with the torque-derived inflow/outflow as function of radius, and are thus not a reliable tracer of real in/outflow.
\item No correlation is found between the candidate radial flow regions from our kinematic study and the inflow/outflow caused by gravitational torques. This might be explained by the fact that the non-circular terms from our residual velocity field are predominantly caused by elliptical motions. 
\item The change in flow direction of the gas caused by gravitational torques can be used to determine the CR radius similarly to other methods (Canzian, Stellar Density-Potential-Phase-Shift, bar length). In particular this method may be applicable in case that several pattern speeds combine. 
\item For the very center (0-0.1~kpc) no significant gas transport is found towards the AGN due to gravitational torques. Thus, other mechanisms such as viscous torques must be important for fueling the AGN at this scale.
\end{itemize}

\acknowledgments
A special thanks goes to Anthony Baillard for providing and assisting us with the tool NFIGI for the removal of foreground stars.
We are grateful to the National Radio Astronomy Observatory (NRAO) for their support during this project.
The NRAO is operated by Associated Universities, Inc., under cooperative agreement with the National Science Foundation. This work is also based on observations carried out with the IRAM Plateau de Bure Interferometer. IRAM is supported by INSU/CNRS (France), MPG (Germany) and IGN (Spain). CGM is grateful for financial support from the Royal Society and Research Councils U.K. SH is supported by the German DFG under grant number SCHI 536/2-1.


\clearpage


\begin{table}
\begin{footnotesize}
\caption{Sample Overview}
\begin{tabular}{lcclcrcll}
\tableline\tableline
Name & RA & DEC & Hubble & AGN& AGN & $v_{sys}$ & Dist \\
       & (J2000) & (J2000) & Type & HFS & Kewley & [km~s$^{-1}$]&[Mpc]\\ 
\tableline 
NGC~3368& 10 46 46 & +11 49 12 & SB(rs)ab  & L2 & L &  897 & 8.1 \\
NGC~3627& 11 20 15 & +12 59 30 & SB(s)b   & L/S2 & S &  727 & 6.6 \\
NGC~4321& 12 22 55 & +15 49 21 & SB(s)bc  & T2 & L & 1571 & 16.8 \\
NGC~4736& 12 50 53 & +41 07 14 & (R)S(r)ab   & L2 & L &  308 & 4.3 \\
NGC~5248& 13 37 32 & +08 53 08 & (R)SB(rs)bc  & H & H & 1153 & 15.0 \\
NGC~6951& 20 37 14 & +66 06 20 & SB(rs)bc & S2 & S & 1424 & 24.1 \\
NGC~7217& 22 07 52 & +31 21 33 & (R)S(r)ab  & L2 & L &  952 & 16.0 \\
\tableline
\end{tabular}
\tablecomments{Summary of the properties of our 7 galaxies. Listed are only parameters from LEDA and NED. The AGN classification listed in column (6) is taken from Ho, Fillipenko \& Sargent (1997): S - Seyfert, L - LINER, T - transition object, H - H~II galaxy and NED. In addition, our AGN classification \citep{Haa08} in Seyfert and LINER  following the method of Kewley et al (\citeyear{Kew06}) is given in column (7). The velocities listed in column (8) are the assumed systemic velocities.}
\label{tab_overview}
\end{footnotesize}
\end{table}

\begin{table}
\caption{Parameters of the VLA HI data}
\begin{tabular}{lrrr}
\tableline\tableline
Name   & Resolution & Resolution & FWZI.\\
 & Beam [$\arcsec$] & [kpc] & [km s$^{-1}$]\\ 
\tableline 
NGC~3368&6.4 $\times$ 5.3& 0.23 &399\\
NGC~3627&5.3 $\times$ 5.1& 0.17 &431\\
NGC~4321&7.6 $\times$ 6.5& 0.57 &291\\
NGC~4736&5.6 $\times$ 5.2& 0.11 &284\\
NGC~5248&5.9 $\times$ 5.4& 0.41 &311\\
NGC~6951&14.4 $\times$ 12.5& 1.57 &338\\
NGC~7217&7.9 $\times$ 5.3& 0.50 &326\\
\tableline
\end{tabular}
\tablecomments{Overview of the spatial resolution of our VLA observations. The velocity resolution is $\sim$10.4 km s$^{-1}$ for all galaxies. In addition the Full Widths at Zero Intensity (FWZI) of the HI lines are listed for all galaxies.}
\label{tab_obs}
\end{table}

\begin{table}
\caption{Parameters of the CO data}
\begin{tabular}{lrrr}
\tableline\tableline
Name   & CO 1-0 & CO 2-1 & Origin\\
 & Beam [$\arcsec$] & Beam [$\arcsec$] & \\ 
\tableline 
NGC~3368 & 2.5 $\times$ 1.4 & 1.0 $\times$ 0.67 & (PI. Schinnerer)\\
NGC~3627 & 1.9 $\times$ 1.2 & 0.9 $\times$ 0.6 & (NUGA)\\
NGC~4321 & 2.2 $\times$ 1.3 & - & \citep{Gar05}\\
NGC~4736 & 2.1 $\times$ 1.7 & -& (PI. Schinnerer)\\ 
NGC~5248 & 6.9 $\times$ 5.8 & -  & \citep{Hel03} \\
NGC~6951 & 2.6 $\times$ 1.7 & 0.6 $\times$ 0.5 & \citep{Gar05}\\
NGC~7217 & 2.4 $\times$ 1.9 & - & \citep{Com04} \\
\tableline
\end{tabular}
\tablecomments{Overview of the angular resolution of our CO observations using the transition J=1-0 and J=2-1 of $^{12}$CO. The velocity resolution is $\sim$5.2 km s$^{-1}$ for all galaxies, except NGC~5248 which has a velocity resolution of $\sim$10 km s$^{-1}$. In addition, the origin of the data is listed.}
\label{tab_COobs}
\end{table}

\begin{table}
\begin{footnotesize}
\caption{NIR data}
\begin{tabular}{lrrrrl}
\tableline\tableline
Name   & Instrument & P.I.; Proposal ID & Instrument &  P.I.; Program ID\\
\tableline 
NGC~3368& NICMOS F160w & J. Mulchaey; 7330& IRAC CH1 & G. Fazio; 69\\
NGC~3627& NICMOS F160w & J. Mulchaey; 7330& IRAC CH1 & R. Kennicutt; 159\\
NGC~4321& K-Band & \cite{Kna03} & IRAC CH1 & R. Kennicutt; 159\\
NGC~4736& NICMOS F160w & R. Kennicutt; 9360& IRAC CH1 & R. Kennicutt; 159\\
NGC~5248& NICMOS F160w & D. Maoz; 7879 & IRAC CH1 & G. Fazio; 69\\
NGC~6951& NICMOS F160w & J. Mulchaey; 7330 & K-Band & \cite{Mul97}\\
NGC~7217& NICMOS F160w & M. Stiavelli; 7331& H-Band & \cite{Esk02}\\
\tableline
\end{tabular}
\tablecomments{Overview of all NIR images used.}
\label{tab_NIR}
\end{footnotesize}
\end{table}

\begin{table}
\caption{Disk and Kinematic Parameters}
\begin{tabular}{lrrrrr}
\tableline\tableline
Source   & x-offset& y-offset & $v_{sys}$ & PA & $i$ \\
         &  [\arcsec] & [\arcsec]  &  [km s$^{-1}$] &  [$\degr$] & [$\degr$] \\ 
\tableline
NGC~3368 &2.2 $\pm$ 0.7 &-2.1 $\pm$ 0.9 &895.5 $\pm$ 0.3 &168.5 $\pm$ 0.5 &56.5 $\pm$ 0.5 \\
NGC~3627 &-0.2 $\pm$ 1.0 &1.1 $\pm$ 2.1 &720.3 $\pm$ 2.2 &174.2 $\pm$ 0.7 &62.0 $\pm$ 2.2\\
NGC~4321 &1.5 $\pm$ 0.5 &0.4 $\pm$ 0.5 &1577.5 $\pm$ 0.4 &152.1 $\pm$ 0.4 &33.0 $\pm$ 1.5\\
NGC~4736 &0.4 $\pm$ 0.3 &0.5 $\pm$ 0.4 &309.0 $\pm$ 0.5 &299.0 $\pm$ 0.9 &34.5 $\pm$ 1.2\\
NGC~5248 &1.6 $\pm$ 0.6 &-1.1 $\pm$ 0.6 &1151.3 $\pm$ 0.3 &113.2 $\pm$ 0.5 &44.6 $\pm$ 0.9\\
NGC~6951 &1.5 $\pm$ 1.0 &-1.7 $\pm$ 1.0 &1424.6 $\pm$ 0.8 &138.1 $\pm$ 0.9 &46.2 $\pm$ 1.1 \\
NGC~7217 &-0.3 $\pm$ 0.5 &0.3 $\pm$ 0.4 &951.8 $\pm$ 0.2 &265.3 $\pm$ 2.0 &35.1 $\pm$ 0.6\\
\tableline
\end{tabular}
\tablecomments{Overview of the disk and kinematic parameters derived from the observed velocity field by fitting tilted-rings to the velocity field: The offset of the dynamical center of the HI disk to the photometric centers from NED, the systemic velocity $v_{sys}$, the position angle PA and the inclination $i$ are listed for all galaxies. Note that the errors are statistical errors and are interdependent for each galaxy.}
\label{tab_par}
\end{table}

\begin{table}
\caption{Kinematic Analysis}
\begin{tabular}{lcc}
\tableline\tableline
Source  & \multicolumn{2}{c}{$\langle v_{nc}/v_{rot} \rangle$} \\
        &             CO      & HI     \\
\tableline 
NGC~3368  &0.132 $\pm$ 0.211 &0.071 $\pm$ 0.120 \\
NGC~3627  &0.832 $\pm$ 2.571 &0.120 $\pm$ 0.084 \\
NGC~4321  &0.330 $\pm$ 0.867 &0.068 $\pm$ 0.064\\
NGC~4736  &0.221 $\pm$ 0.153 &0.130 $\pm$ 0.300\\
NGC~5248  &0.115 $\pm$ 0.042 &0.098 $\pm$ 0.063\\
NGC~6951  &0.137 $\pm$ 0.029 &0.066 $\pm$ 0.293\\
NGC~7217  &0.071 $\pm$ 0.093 &0.054 $\pm$ 0.056\\
\tableline
\textbf{Mean} & 0.26 & 0.09\\
\tableline
\end{tabular}
\tablecomments{Overview of the ratio of non-circular motion to rotation velocity $v_{nc}/v_{rot}$  with the standard deviation of the mean value for CO and HI.}
\label{tab_kin}
\end{table}

\begin{table}
\caption{Error estimation}
\begin{tabular}{lcclccrcll}
\tableline\tableline
Parameter & d$\tau$ (CO) & d$\tau$ (HI)\\
\tableline 
Position Angle (dPA=2$\degr$) & 4\% & $<$1\% \\
Inclination (d$i$=2$\degr$) & 24\% &  $<$1\% \\
Shift of IR image (0.2$\arcsec$)& 10-30\% & $\sim$13\%;\\
Shift of gas image (0.3$\arcsec$) & 5-20\% & 5-20\%\\
Background subtraction & $<$5\% & $<$1\% \\
Bulge-nobulge subtraction & $<2$\% & 3-10\% \\
Rot vel (d$v$=15~km~s$^{-1}$) & 20\%  & 10\%  \\
\tableline
\textbf{Sum} & 60-80\% & 13-70\% \\
\textbf{Typical} & 67\% & 30\% \\
\tableline
\end{tabular}
\tablecomments{Estimation of systematic errors in the radial torques for CO and HI of a typical galaxy from our sample, namely NGC~6951. For this exercise, the torque results using the best input parameters have been compared with those where typical errors of the input parameters were applied. Note that the errors have a radial dependency (see Fig.~\ref{fig_test6951})}
\label{tab_errors}
\end{table}

\begin{rotate}
\begin{table}
\begin{footnotesize}
\caption{Resonances}
\begin{tabular}{lrrrrrrrrl}
\tableline\tableline
Source  & $CR_{Torque}$ & $R_{Bar}$ & $CR_C$ & $CR_{PS}$ & $\Omega_{Bar}$ & ILR & UHR & OLR & Comments  \\
 & [kpc] & [kpc] & [kpc] & [kpc] & [km~s$^{-1}$~kpc$^{-1}$]& [kpc] & [kpc] & [kpc] &\\
\tableline 
NGC~3368 & 4.5 & 3.2 & 2.5-3.5& \textbf{3.5}& 54 & - & 2.5 & 7.4& Outer Bar\\
         &     &    &  7     & \textbf{7.0}&   &    &     &       & Outer Spiral\\    
NGC~3627 & 2.7 & 1.5 & 2 & \textbf{2.4} & 79 & - & 0.4 & 3.7 & Outer Bar\\
NGC~4321 &     & 5 & 8.5& \textbf{4.8} & 39 & - & 1.7 &  13.0 &Bar\\
         & 11.7&    &     &\textbf{11.2} &    &     &    &   & Outer Spiral\\    
NGC~4736 &  -  &  - & 5.5-7 &  &  &  & & & Outer Oval \\
NGC~5248 &     &    & 1.7-2.0& \textbf{3.6} & 41 & 2.3 & 2.7 & 8.9 & Inner Spiral/Bar \\
         &9.5-11& 7& 7.5-12 & \textbf{11.8}&    &     &     &     &Outer Spiral/Bar\\    
NGC~6951 &    &7 & 5-7  & \textbf{8.7} & 23 & (1.6, 2.5) &3.2 &20.0 & Bar  \\
         &  11 &  &      &  \textbf{16.2}  &   &    &     &       & Outer Resonance\\ 
         &     &  & 20-23 &  22.1  &    &    &     &       & Outer Resonance\\   
NGC~7217 & 5.5-7 & - & -  & - & -   & -  &- & - & No Bar or Spiral \\
\tableline
\end{tabular}
\tablecomments{Overview of the dynamical resonances derived from different methods: the CR radius from our gravity torque study (column 2), the radius of the bar (column 3), the CR using the method of \cite{Can93} (column 4), the CR on the basis of the Phase-Shift method of \cite{Zha07} (column 5) and the associated angular speed of the bar $\Omega_{Bar}$ (column 6), the Inner Lindblad Resonance (ILR) at $\Omega_{Bar}=\Omega-\kappa/2$ (column 7), the Ultra Harmonic Resonance (UHR) at $\Omega_{Bar}=\Omega-\kappa/4$ (column 8), and the Outer Lindblad Resonance (OLR) at $\Omega_{Bar}=\Omega+\kappa/2$ (column 9) of the bar. All radii are deprojected. The CR values printed in boldface are taken for further interpretation of our torque results as a function of the CR locations of spiral arms and bars.}
\label{tab_bar}
\end{footnotesize}
\end{table}
\end{rotate}

\begin{table}
\begin{footnotesize}
\caption{Gravitational Torque Analysis}
\begin{tabular}{l|rr|rr|rr|rr}
\tableline\tableline
Source  & \multicolumn{2}{c}{$\langle dL/L \rangle$} \vline & \multicolumn{2}{c}{$\langle \vert dL/L \vert \rangle$} \vline & $dM_{HI}/dt$ & $dM_{mol}/dt$ & $R_{max,HI}$ & $R_{max,CO}$\\
 & HI & CO & HI & CO & [M$_{\odot}$~yr$^{-1}]$ &  [M$_{\odot}$~yr$^{-1}]$ & [kpc] & [kpc]\\
\tableline 
NGC~3368 &0.02 &-0.50 &0.12 &0.55 &0.04 &-11.11 & 13.0 & 0.35 \\
NGC~3627 &-0.00 &0.16 &0.26 &0.24 &-0.10 &50.13 & 6.0 & 0.75 \\
NGC~4321 &0.01 &0.02 &0.10 &0.10 &-0.18 &1.03 & 18.0 & 1.2 \\
NGC~4736 &0.02 &0.01 &0.03 &0.09 &0.06 &-0.01 & 8.5 & 0.62 \\
NGC~5248 &0.06 &-0.02 &0.13 &0.02 &0.49 &-22.96 & 20.5 & 0.8 \\
NGC~6951 &-0.04 &-0.05 &0.09 &0.10 &-0.16 &-4.29 & 29.0 & 1.0 \\
NGC~7217 &-0.01 &-0.02 &0.03 &0.03 &-0.01 &-2.37 & 15.2 & 1.4  \\
\tableline
\end{tabular}
\tablecomments{Overview of the transfered angular momentum and averaged transfered gas masses over the entire gas disk for the center (molecular gas traced by CO) and the outer disk (atomic gas traced by HI). A positive (negative) sign corresponds to a net outflow (inflow). The averaged transfered angular momentum $\langle {dL/L} \rangle$ over the entire gas disk is listed in column 2 and 3 and the mean value of the absolute angular momentum transfer within the disk in column 3 and 4.
The net mass inflow/outflow rates d$M(R)$/(d$t$) (in units of $M_{\odot} yr^{-1}~pc^{-1}$) out to the maximum disk radius (column 8,9) are listed in column 6 (7) for the atomic (molecular) gas.}
\label{tab_torque}
\end{footnotesize}
\end{table}

\clearpage


\begin{figure}[ht]
\begin{center}
\begin{minipage}[c][15cm][t]{8cm}
\includegraphics[scale=0.42]{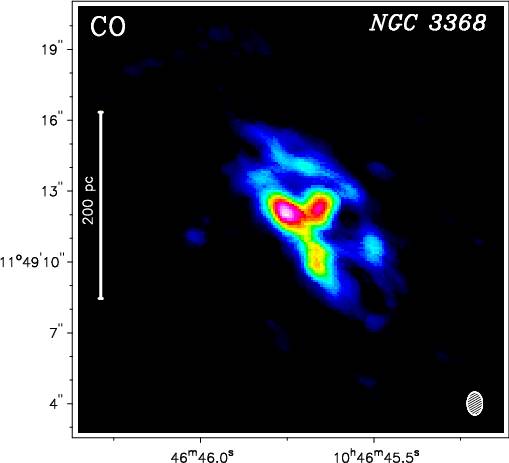}
\includegraphics[scale=0.58]{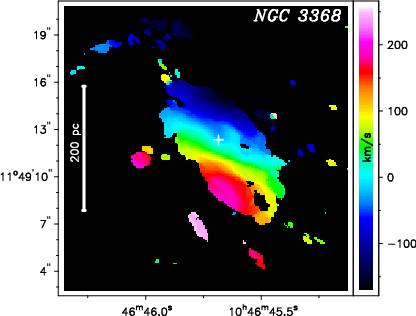}
\end{minipage}
\begin{minipage}[c][15cm][t]{1cm}
\end{minipage}
\begin{minipage}[c][15cm][t]{8cm}
\includegraphics[scale=0.42]{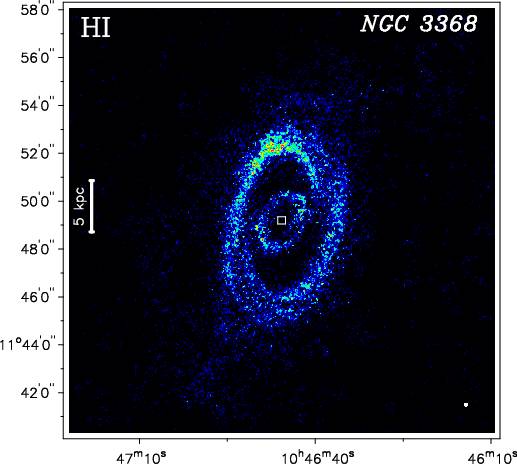}
\includegraphics[scale=0.58]{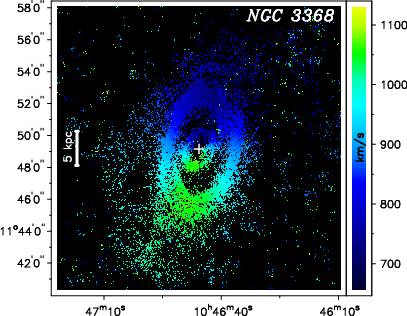}
\end{minipage} 
\end{center}
\caption{\footnotesize
Overview of the intensity maps (top panels) and intensity-weighted velocity fields (bottom panels) for the molecular (left panels) and atomic gas (right panels). The beam is plotted in the bottom left corner and a scale length of 1~kpc (200~pc) is indicated as bar for the HI (CO) maps at the left. Only for NGC~5248 the bar indicates a scale length of 1~kpc for the CO map. The dynamical center is indicated by a cross in the velocity fields (bottom panels). Colorbars are centered at $v_{center}=0$ for the molecular gas velocity fields (except for the BIMA-SONG observation of NGC~5248) or the systemic velocity ($v_{center}=v_{sys}$) for the atomic gas velocity fields. The square in the HI map indicates the FOV of the CO map. The molecular gas has been observed in the transition of the J=1-0 line of $^{12}$CO for almost all galaxies. Only for NGC~7217 the molecular gas observation in the J=2-1 line of $^{12}$CO is shown.}  
\label{fig_kin}
\end{figure}

\begin{figure}[ht]
\begin{center}
\begin{minipage}[c][15cm][t]{8cm}
\includegraphics[scale=0.42]{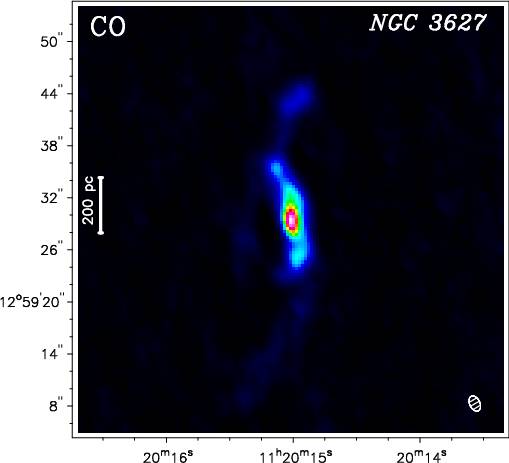}
\includegraphics[scale=0.58]{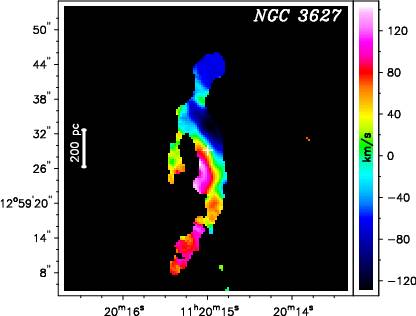}
\end{minipage}
\begin{minipage}[c][15cm][t]{1cm}
\end{minipage}
\begin{minipage}[c][15cm][t]{8cm}
\includegraphics[scale=0.42]{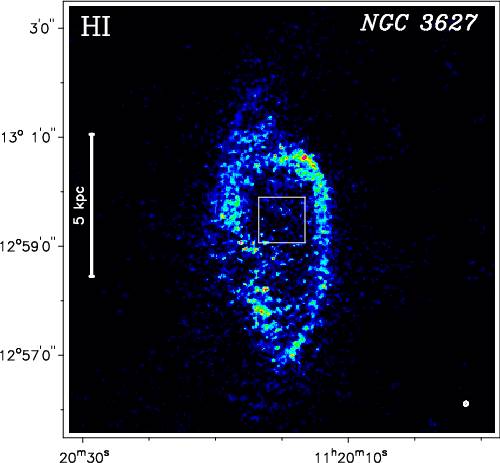}
\includegraphics[scale=0.58]{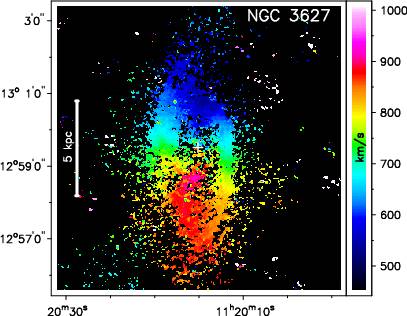}
\end{minipage} 
\end{center}
\figurenum{\ref{fig_kin}}
\caption{(Continued)}  
\end{figure}

\begin{figure}[ht]
\begin{center}
\begin{minipage}[c][15cm][t]{8cm}
\includegraphics[scale=0.42]{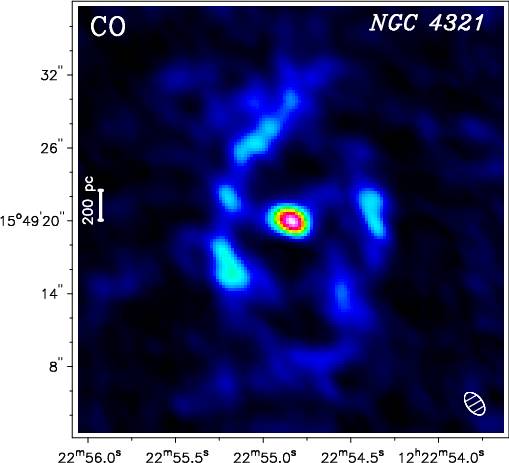}
\includegraphics[scale=0.58]{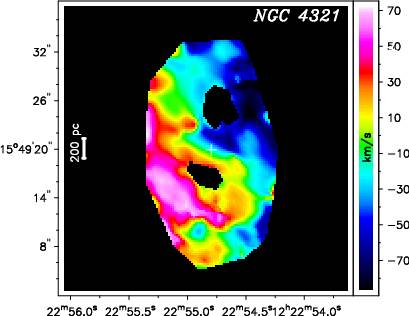}
\end{minipage}
\begin{minipage}[c][15cm][t]{1cm}
\end{minipage}
\begin{minipage}[c][15cm][t]{8cm}
\includegraphics[scale=0.42]{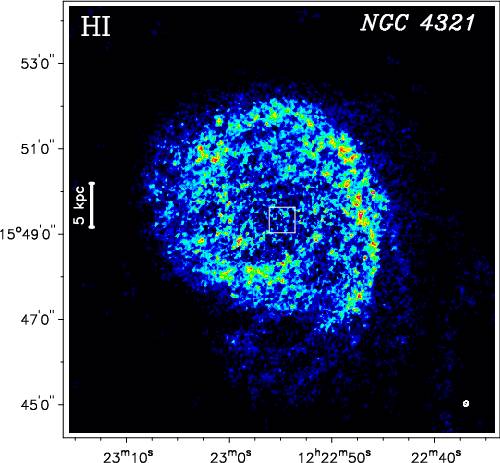}
\includegraphics[scale=0.58]{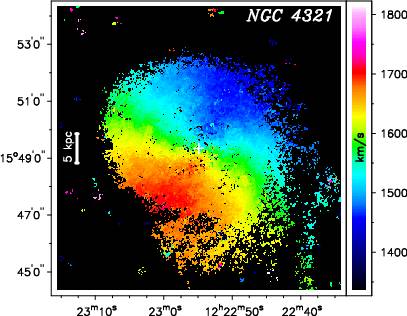}
\end{minipage} 
\end{center}
\figurenum{\ref{fig_kin}}
\caption{(Continued)}  
\end{figure}

\begin{figure}[ht]
\begin{center}
\begin{minipage}[c][15cm][t]{8cm}
\includegraphics[scale=0.42]{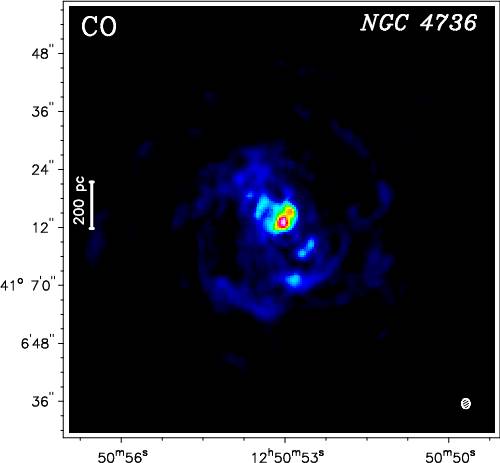}
\includegraphics[scale=0.58]{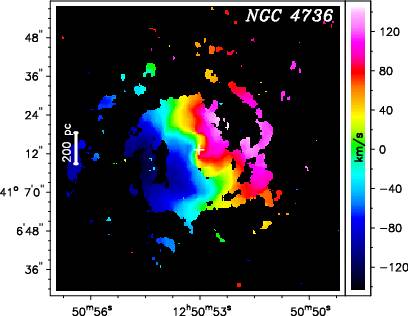}
\end{minipage}
\begin{minipage}[c][15cm][t]{1cm}
\end{minipage}
\begin{minipage}[c][15cm][t]{8cm}
\includegraphics[scale=0.42]{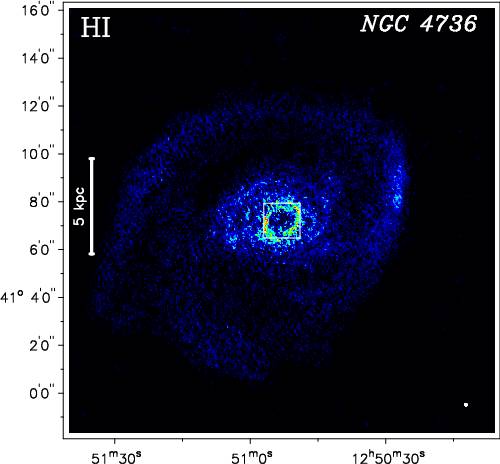}
\includegraphics[scale=0.58]{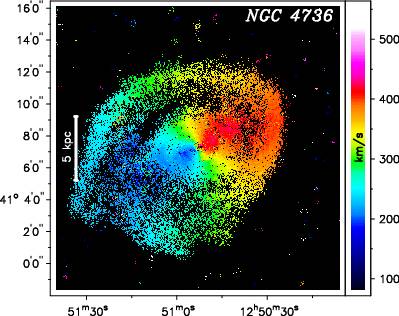}
\end{minipage} 
\end{center}
\figurenum{\ref{fig_kin}}
\caption{(Continued)}  
\end{figure}

\begin{figure}[ht]
\begin{center}
\begin{minipage}[c][15cm][t]{8cm}
\includegraphics[scale=0.42]{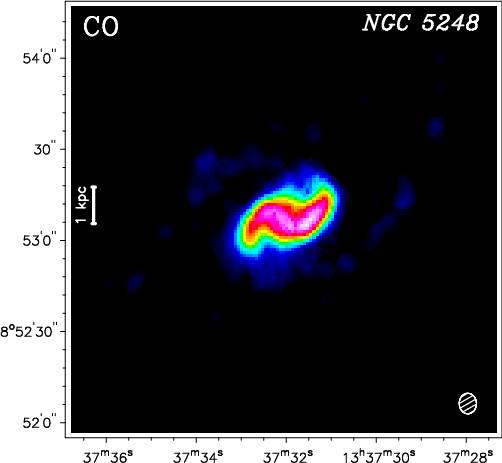}
\includegraphics[scale=0.58]{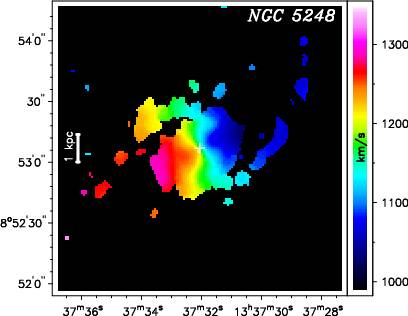}
\end{minipage}
\begin{minipage}[c][15cm][t]{1cm}
\end{minipage}
\begin{minipage}[c][15cm][t]{8cm}
\includegraphics[scale=0.42]{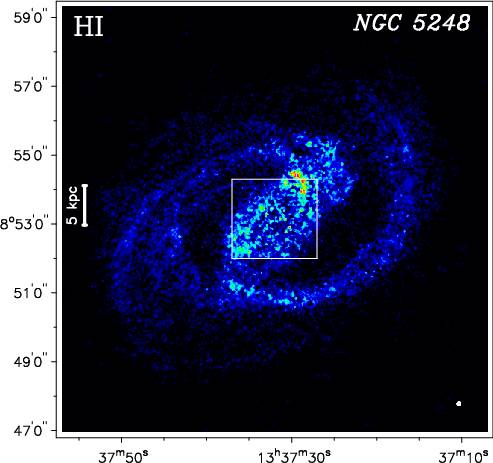}
\includegraphics[scale=0.58]{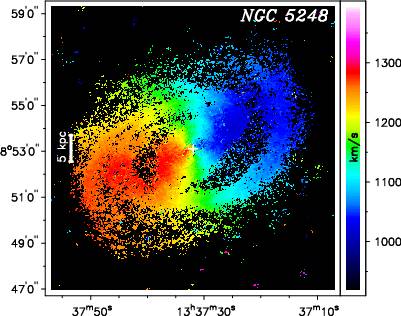}
\end{minipage} 
\end{center}
\figurenum{\ref{fig_kin}}
\caption{(Continued)} 
\end{figure}

\begin{figure}[ht]
\begin{center}
\begin{minipage}[c][15cm][t]{8cm}
\includegraphics[scale=0.42]{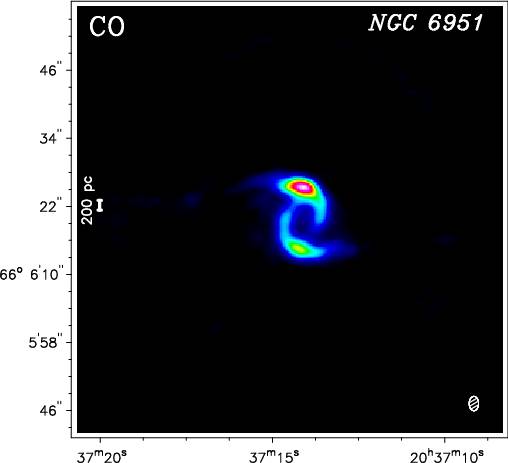}
\includegraphics[scale=0.58]{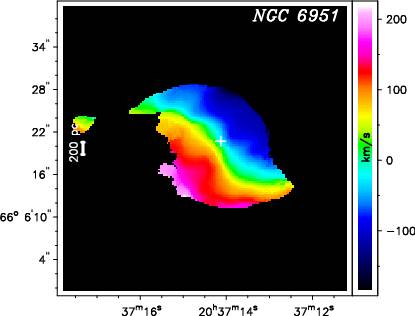}
\end{minipage}
\begin{minipage}[c][15cm][t]{1cm}
\end{minipage}
\begin{minipage}[c][15cm][t]{8cm}
\includegraphics[scale=0.42]{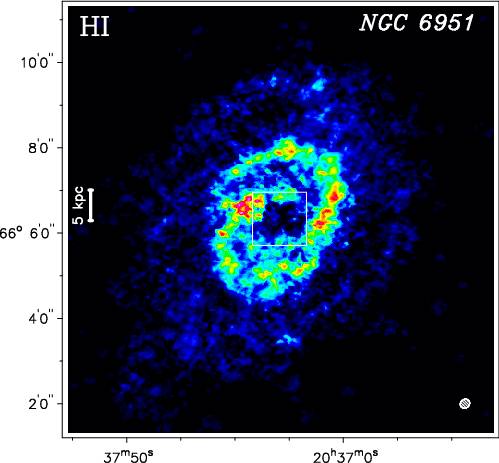}
\includegraphics[scale=0.58]{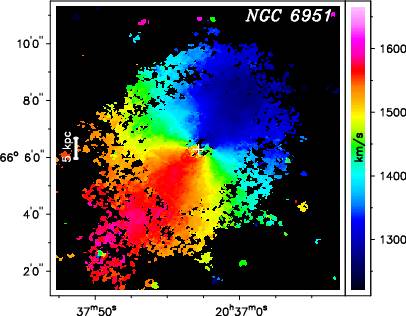}
\end{minipage} 
\end{center}
\figurenum{\ref{fig_kin}}
\caption{(Continued)}  
\end{figure}

\begin{figure}[ht]
\begin{center}
\begin{minipage}[c][15cm][t]{8cm}
\includegraphics[scale=0.42]{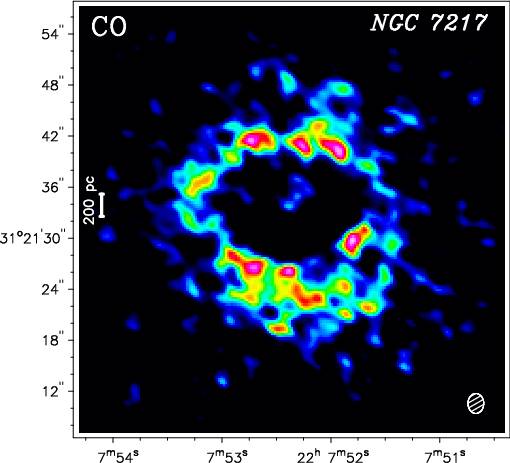}
\includegraphics[scale=0.58]{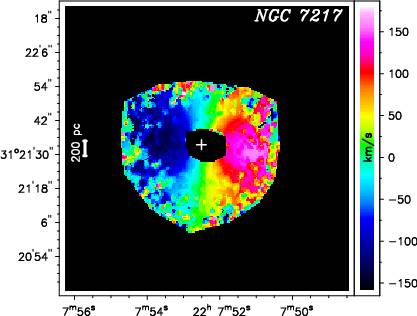}
\end{minipage}
\begin{minipage}[c][15cm][t]{1cm}
\end{minipage}
\begin{minipage}[c][15cm][t]{8cm}
\includegraphics[scale=0.42]{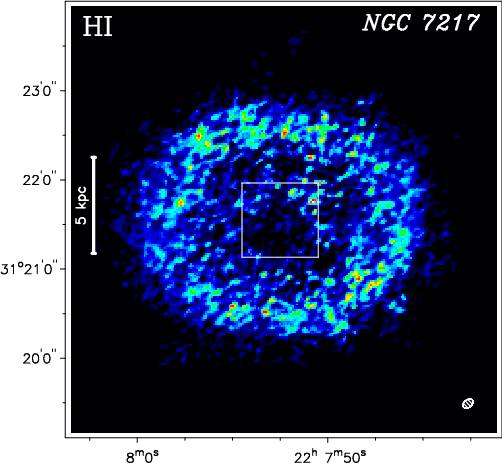}
\includegraphics[scale=0.58]{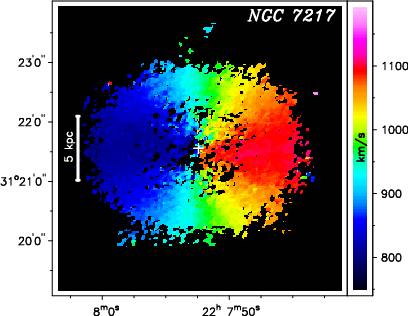}
\end{minipage} 
\end{center}
\figurenum{\ref{fig_kin}}
\caption{(Continued)}  
\end{figure}

\clearpage

\begin{figure}[ht]
\begin{center}
\includegraphics[scale=0.8]{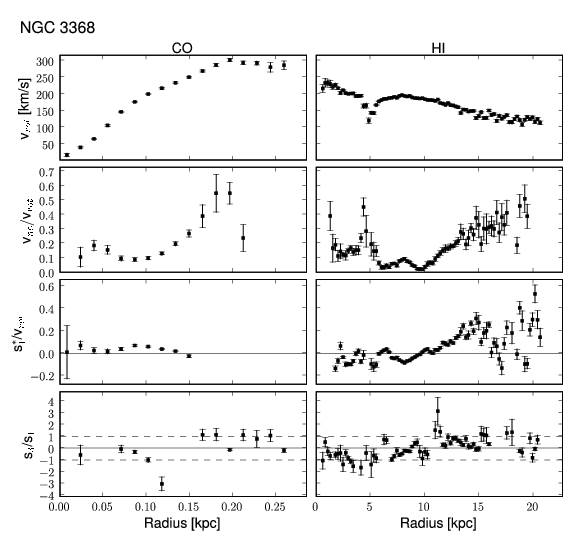}
\end{center}
\caption{\footnotesize Overview of the modeling results based on the CO (left panels) and HI (right panels) velocity fields. From the top to the bottom panels: The rotation curve $v_{rot}$, the ratio of non-circular motion to rotation velocity $v_{nc}/v_{rot}$, the ratio of the amplitude of the first sine-term to the rotation velocity $s_1^*/v_{rot}$, and the ratio of $s_3/s_1$ as a function of radius for each galaxy. Candidate radial gas outflow and inflow are characterized by a positive and negative sign of the $s_1^*/v_{rot}$ term as long as $\vert s_3/s_1 \vert \ll 1$. Data with error bars larger than three (two) times the median of the dataset are clipped for HI (CO). At the CR radius $R_{CR}$ the dominance switches from the $s_1$- to the $s_3$-terms \citep{Can97} as indicated by the horizontal dashed lines at $s_3/s_1=\pm 1$.}
\label{fig_harmonics}
\end{figure}

\begin{figure}[ht]
\begin{center}
\includegraphics[scale=0.8]{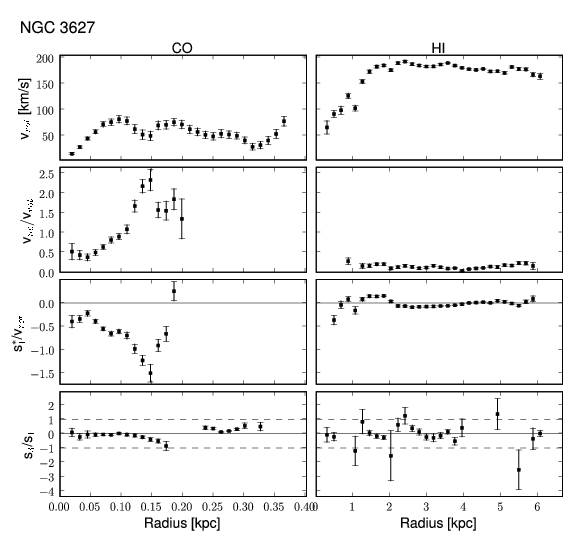}
\end{center}
\figurenum{\ref{fig_harmonics}}
\caption{(Continued)}
\end{figure}

\begin{figure}[ht]
\begin{center}
\includegraphics[scale=0.8]{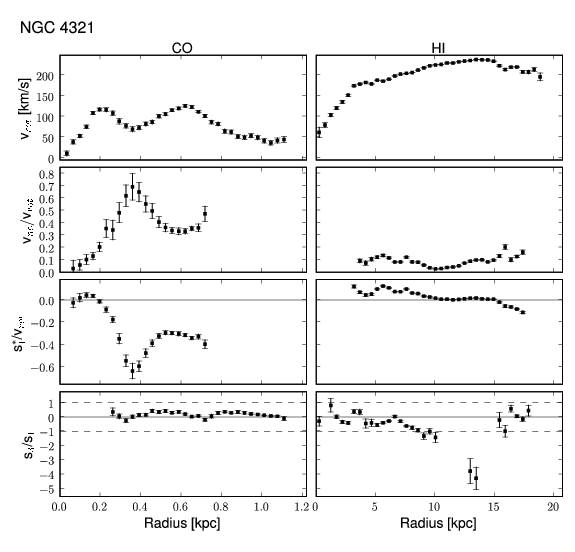}
\end{center}
\figurenum{\ref{fig_harmonics}}
\caption{(Continued)}
\end{figure}

\begin{figure}[ht]
\begin{center}
\includegraphics[scale=0.8]{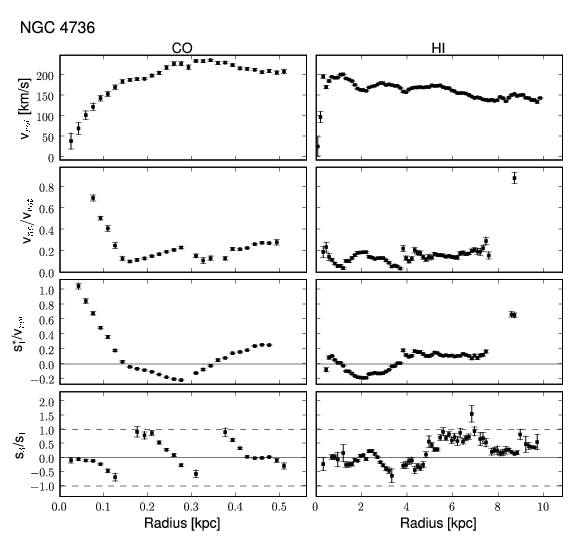}
\end{center}
\figurenum{\ref{fig_harmonics}}
\caption{(Continued)}
\end{figure}

\begin{figure}[ht]
\begin{center}
\includegraphics[scale=0.8]{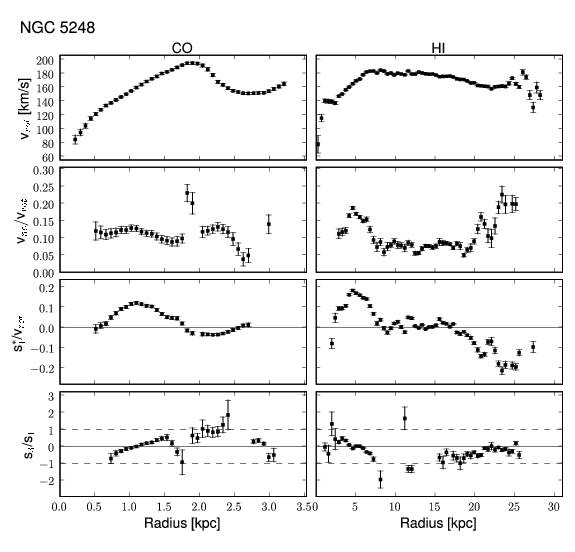}
\end{center}
\figurenum{\ref{fig_harmonics}}
\caption{(Continued)}
\end{figure}

\begin{figure}[ht]
\begin{center}
\includegraphics[scale=0.8]{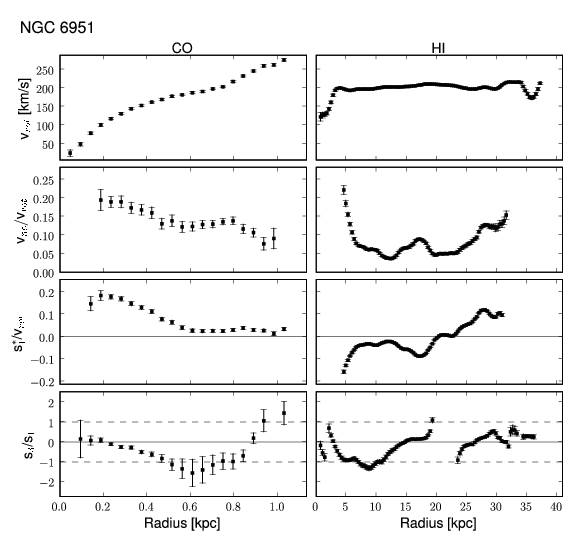}
\end{center}
\figurenum{\ref{fig_harmonics}}
\caption{(Continued)}
\end{figure}

\begin{figure}[ht]
\begin{center}
\includegraphics[scale=0.8]{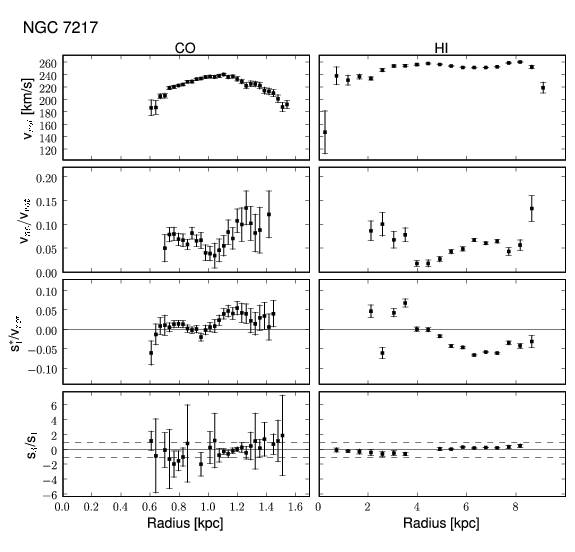}
\end{center}
\figurenum{\ref{fig_harmonics}}
\caption{(Continued)}
\end{figure}

\clearpage

\begin{figure}[ht]
\begin{center}
\includegraphics[scale=0.2]{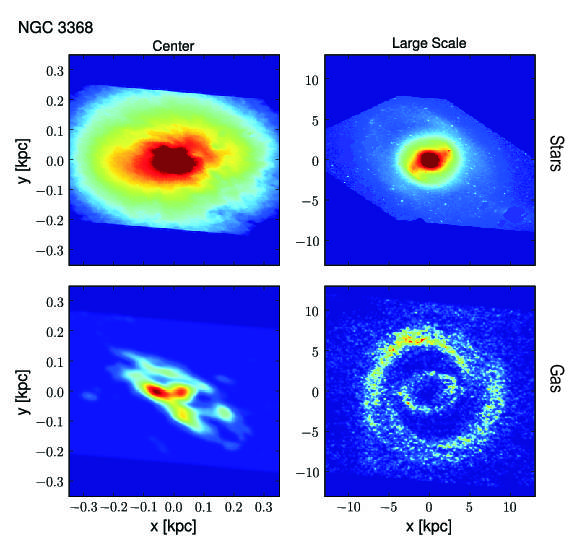}
\includegraphics[scale=0.2]{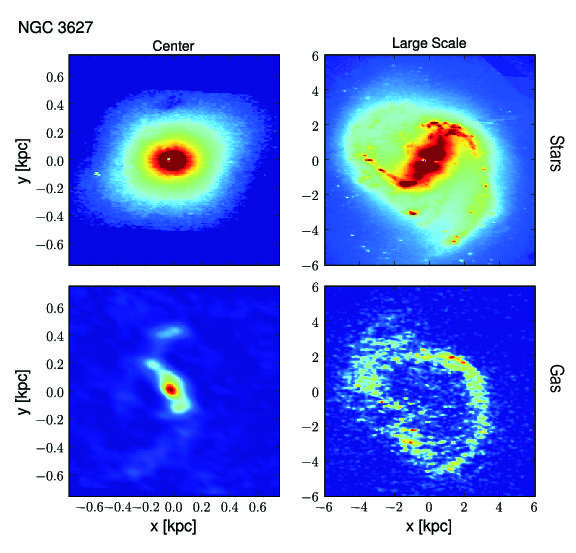}
\end{center}
\caption{\footnotesize Overview of the deprojected stellar (top panels) and gaseous distribution (bottom panels) traced by our CO (left bottom panels) and HI observations (right bottom panels).}
\label{fig_torq_dpj}
\end{figure}

\begin{figure}[ht]
\begin{center}
\includegraphics[scale=0.2]{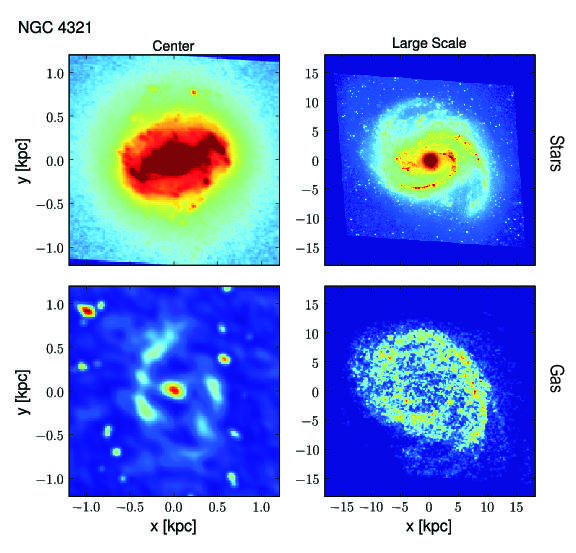}
\includegraphics[scale=0.2]{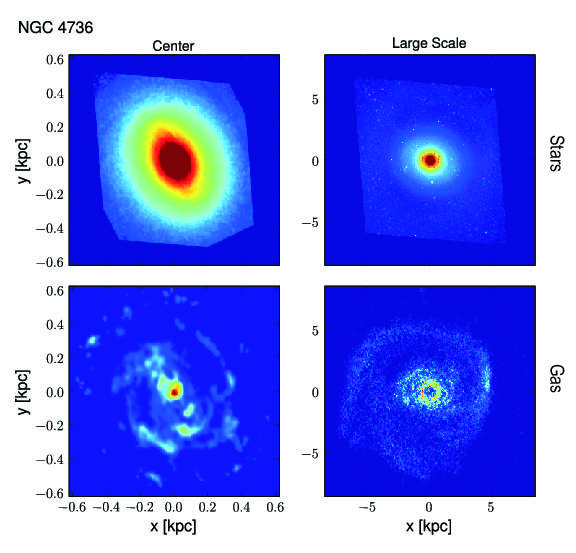}
\end{center}
\figurenum{\ref{fig_torq_dpj}}
\caption{(Continued)}
\end{figure}

\begin{figure}[ht]
\begin{center}
\includegraphics[scale=0.2]{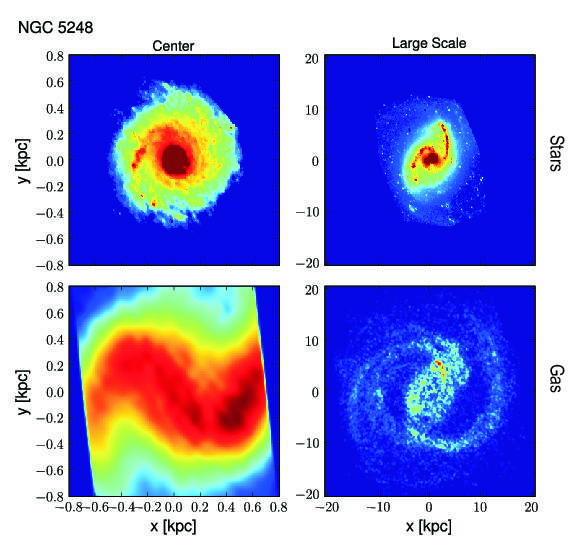}
\includegraphics[scale=0.2]{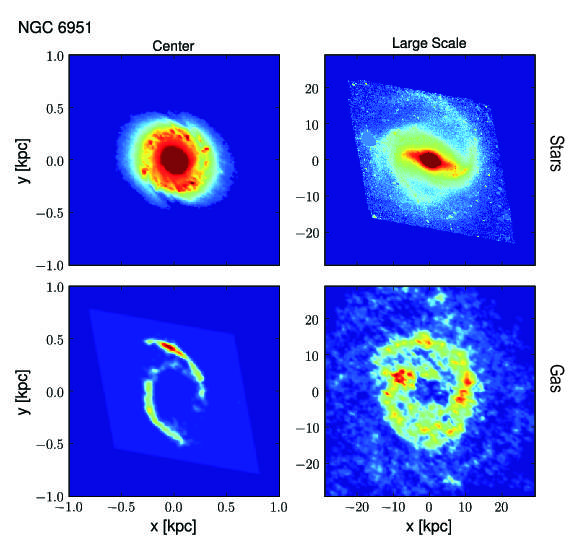}
\end{center}
\figurenum{\ref{fig_torq_dpj}}
\caption{(Continued)}
\end{figure}

\begin{figure}[ht]
\begin{center}
\includegraphics[scale=0.2]{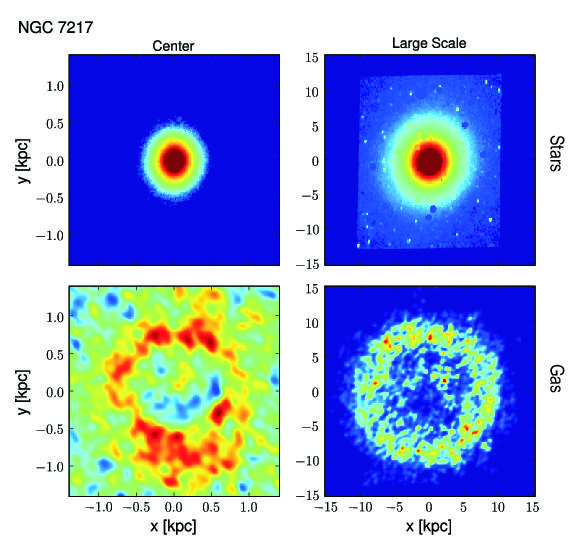}
\end{center}
\figurenum{\ref{fig_torq_dpj}}
\caption{(Continued)}
\end{figure}

\clearpage

\clearpage

\clearpage

\begin{figure}[ht]
\begin{center}
\includegraphics[scale=0.65]{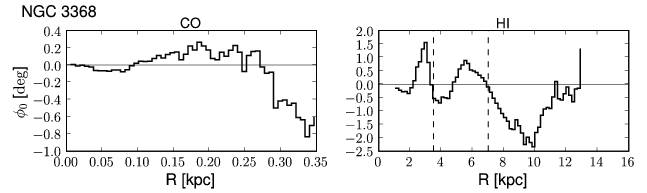}
\includegraphics[scale=0.65]{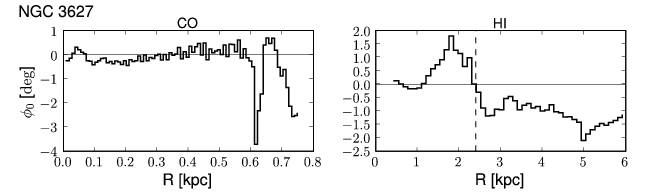}
\includegraphics[scale=0.65]{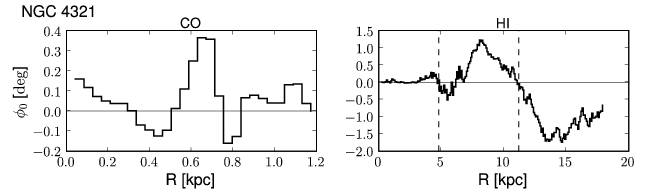}
\includegraphics[scale=0.65]{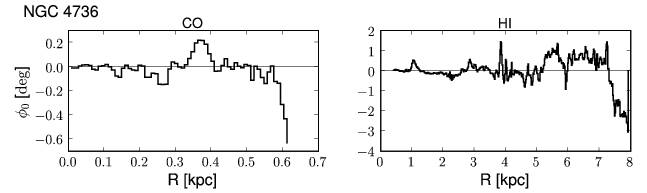}
\end{center}
\caption{\footnotesize Overview of the phaseshift between the potential and density wave patterns for the center (left panels) and the large scale disk (right panels). The CR radii are defined as the positive-to-negative crossings of the phaseshift and indicated by dashed vertical lines.}
\label{fig_phase}
\end{figure}
\begin{figure}[ht]
\begin{center}
\includegraphics[scale=0.65]{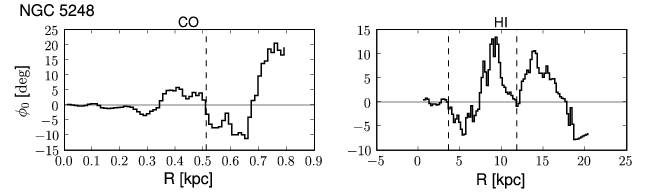}
\includegraphics[scale=0.65]{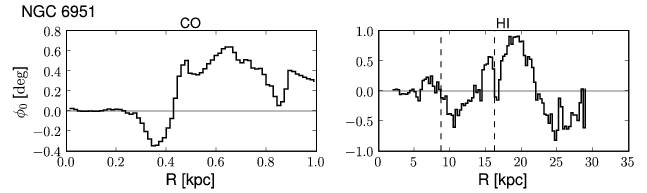}
\includegraphics[scale=0.65]{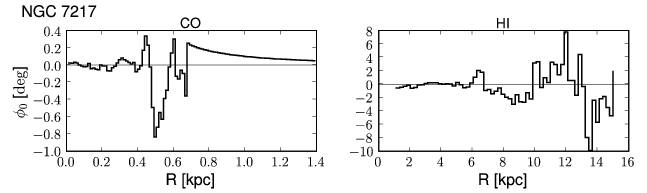}
\end{center}
\figurenum{\ref{fig_phase}}
\caption{(Continued)}
\end{figure}

\clearpage

\begin{figure}[ht]
\begin{center}
\includegraphics[scale=0.7]{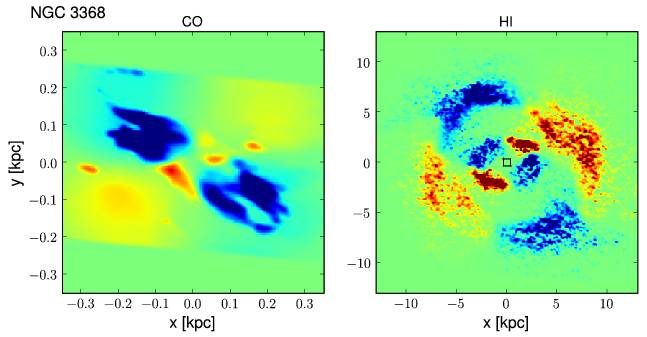}
\includegraphics[scale=0.7]{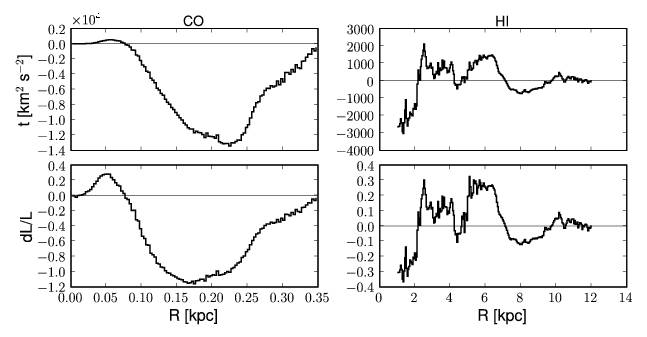}
\end{center}
\caption{\footnotesize Overview of the torque results for the central (CO, left panels) and the large scale disk (HI, right panels). A positive (red colored) or a negative (blue colored) torque corresponds to an outflow or inflow of gas, respectively. The top panels show the gravitational torques weighted with the CO- (left top panel) and HI- (right top panel) gas column density and normalized to their the maximum values. The square in the HI-weighted torque map (right top panel) indicates the FOV of the CO-weighted torque map. The torque per unit mass averaged over azimuth $\tau(R)$ and the fraction of angular momentum d$L/L$ transfered from/to the gas in one rotation are shown in the center and bottom panels, respectively.}
\label{fig_torque}
\end{figure}
\begin{figure}[ht]
\begin{center}
\includegraphics[scale=0.7]{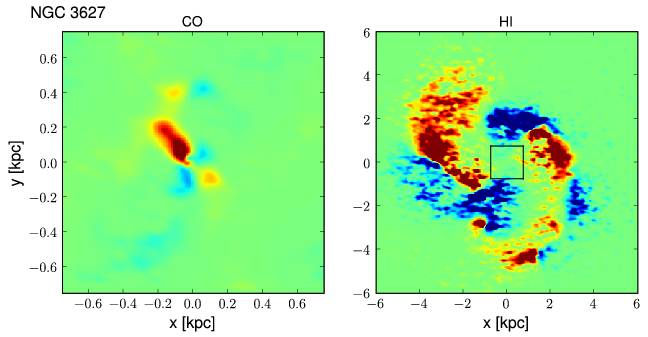}
\includegraphics[scale=0.7]{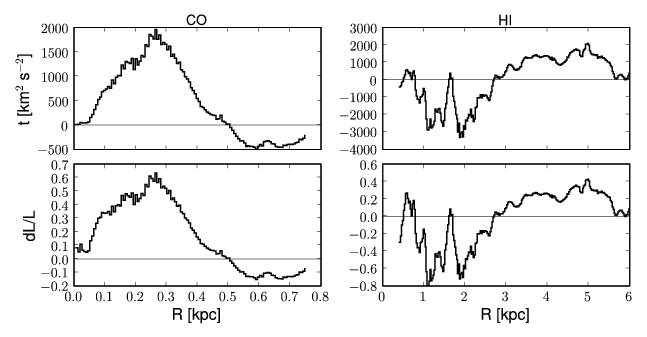}
\end{center}
\figurenum{\ref{fig_torque}}
\caption{(Continued)}
\end{figure}
\begin{figure}[ht]
\begin{center}
\includegraphics[scale=0.7]{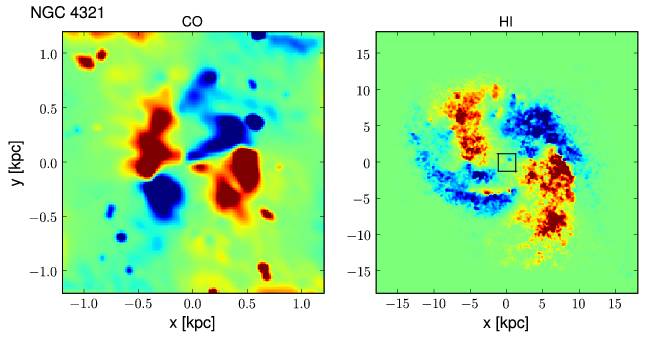}
\includegraphics[scale=0.7]{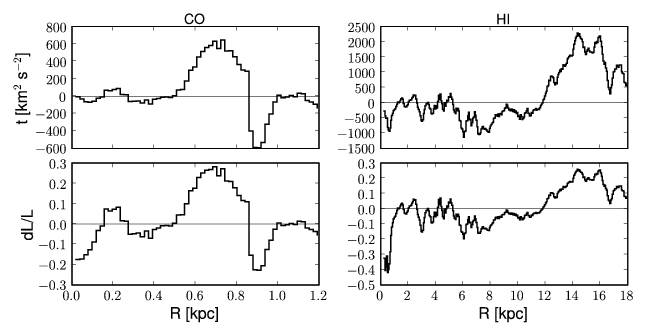}
\end{center}
\figurenum{\ref{fig_torque}}
\caption{(Continued)}
\end{figure}
\begin{figure}[ht]
\begin{center}
\includegraphics[scale=0.7]{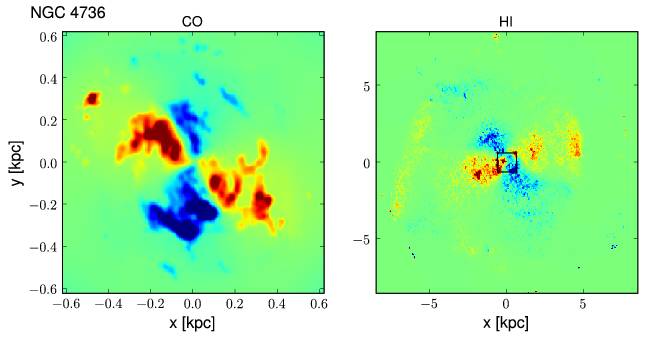}
\includegraphics[scale=0.7]{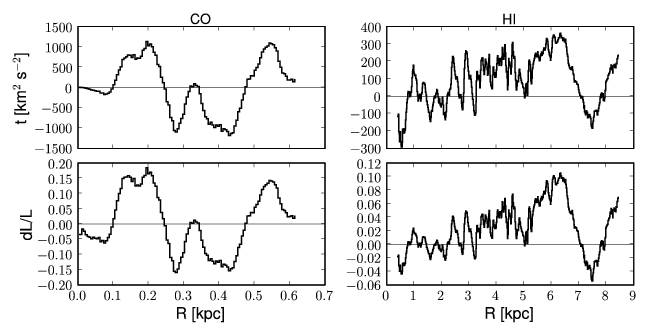}
\end{center}
\figurenum{\ref{fig_torque}}
\caption{(Continued)}
\end{figure}
\begin{figure}[ht]
\begin{center}
\includegraphics[scale=0.7]{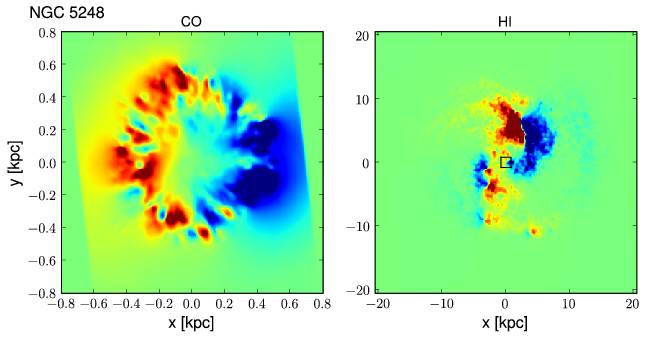}
\includegraphics[scale=0.7]{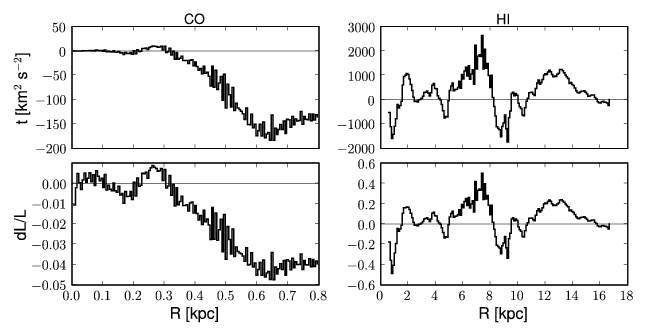}
\end{center}
\figurenum{\ref{fig_torque}}
\caption{(Continued)}
\end{figure}
\begin{figure}[ht]
\begin{center}
\includegraphics[scale=0.7]{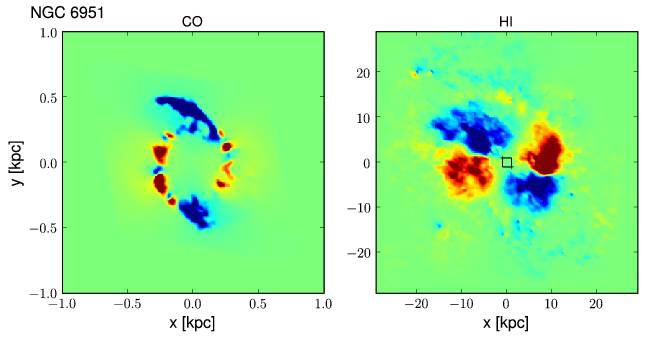}
\includegraphics[scale=0.7]{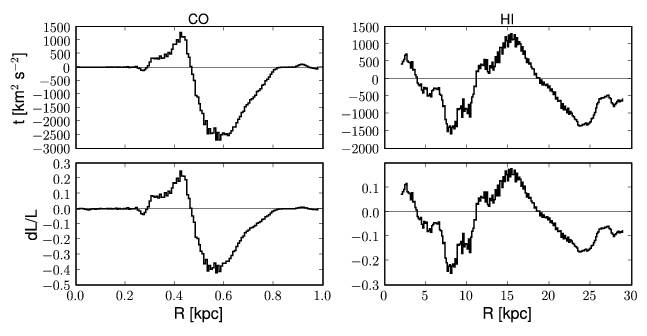}
\end{center}
\figurenum{\ref{fig_torque}}
\caption{(Continued)}
\end{figure}
\begin{figure}[ht]
\begin{center}
\includegraphics[scale=0.7]{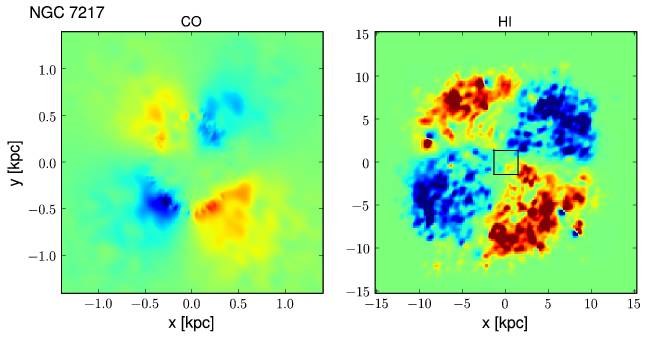}
\includegraphics[scale=0.7]{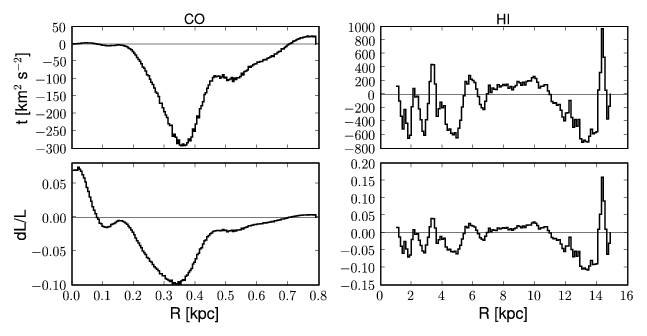}
\end{center}
\figurenum{\ref{fig_torque}}
\caption{(Continued)}
\end{figure}

\clearpage

\begin{figure}[ht]
\begin{center}
\includegraphics[scale=0.65]{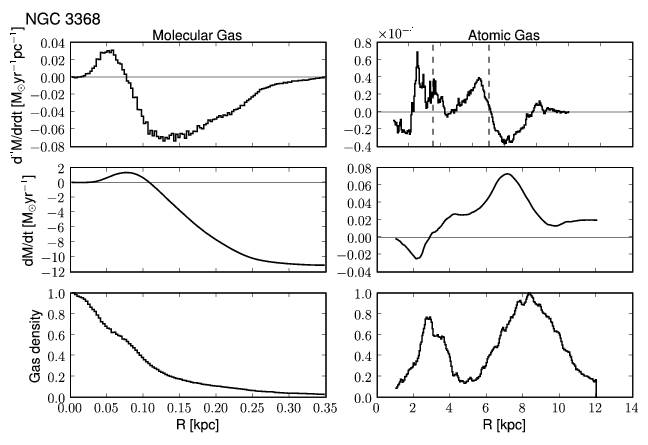}
\end{center}
\caption{ Overview of the mass flow rates for the molecular (left panels) and atomic gas (right panels). The upper panels present the mass flow rate of gas per unit length while the central panels show the cumulative mass flow rate of gas out to a given radius. The molecular and atomic gas densities (bottom panels) are normalized to the maximum density. The vertical dashed line indicates the CR radius of single density wave pattern (bar or spiral), if present, derived from the Stellar-Density-Phase-Shift method (see \S \ref{subsec:torque_phase}).}
\label{fig_torque_mass}
\end{figure}
\begin{figure}[ht]
\begin{center}
\includegraphics[scale=0.65]{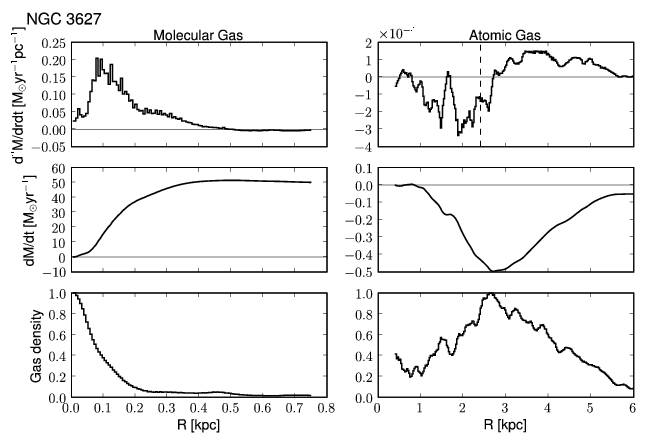}
\includegraphics[scale=0.65]{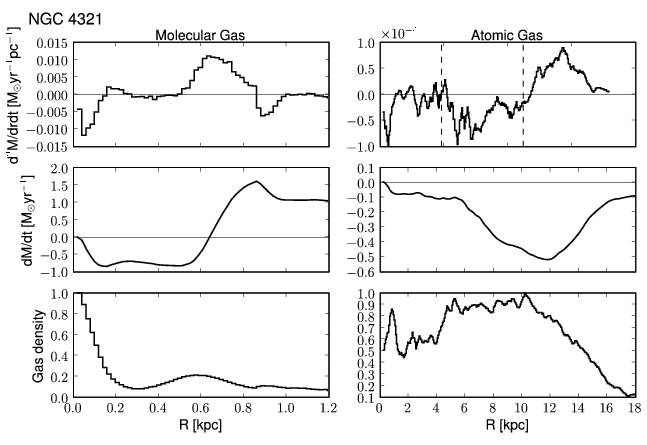}
\end{center}
\figurenum{\ref{fig_torque_mass}}
\caption{(Continued)}
\end{figure}
\begin{figure}[ht]
\begin{center}
\includegraphics[scale=0.65]{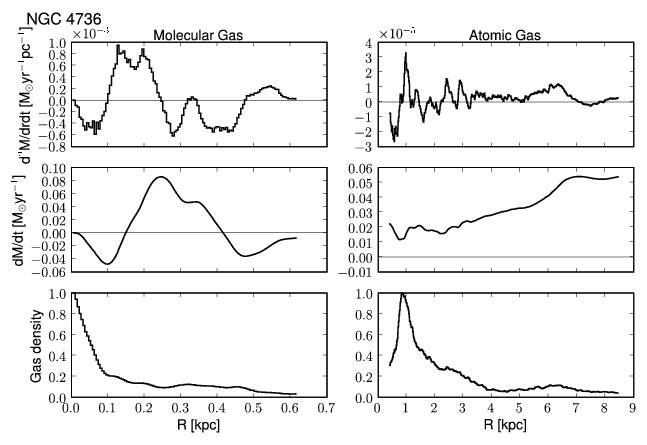}
\includegraphics[scale=0.65]{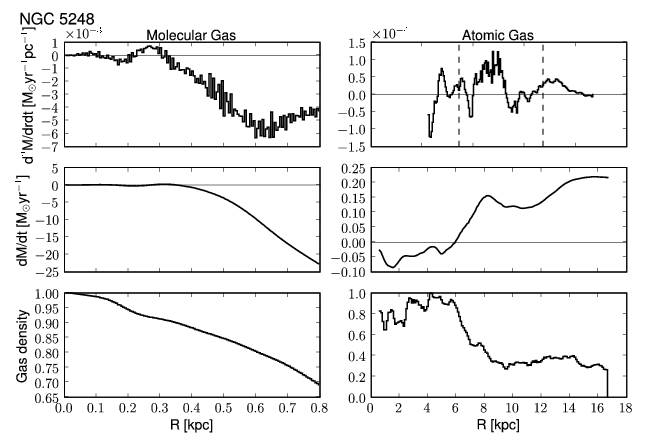}
\end{center}
\figurenum{\ref{fig_torque_mass}}
\caption{(Continued)}
\end{figure}
\begin{figure}[ht]
\begin{center}
\includegraphics[scale=0.65]{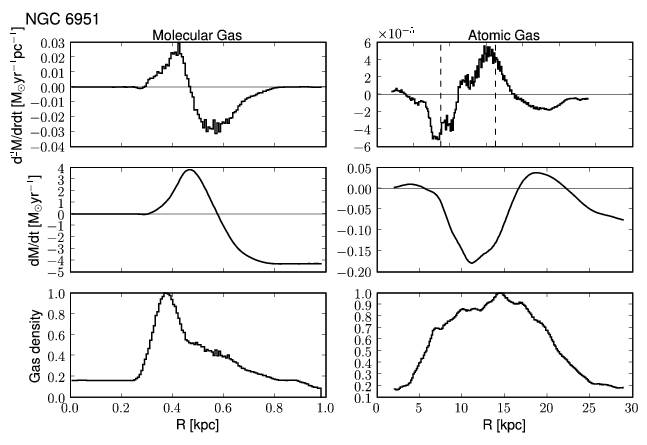}
\includegraphics[scale=0.65]{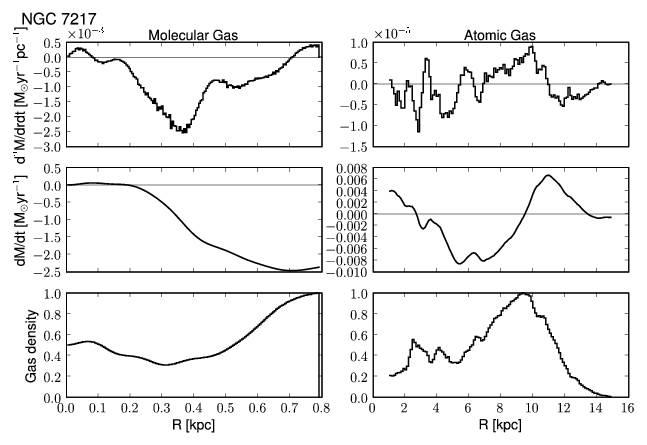}
\end{center}
\figurenum{\ref{fig_torque_mass}}
\caption{(Continued)}
\end{figure}

\clearpage

\begin{figure}[ht]
\begin{center}
\includegraphics[scale=0.62]{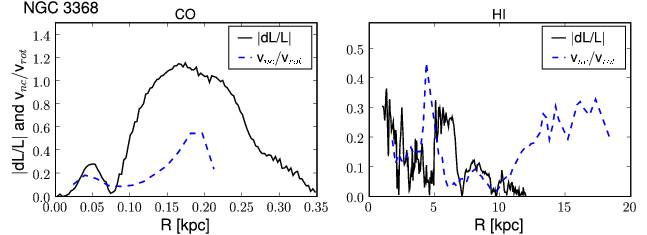}
\includegraphics[scale=0.62]{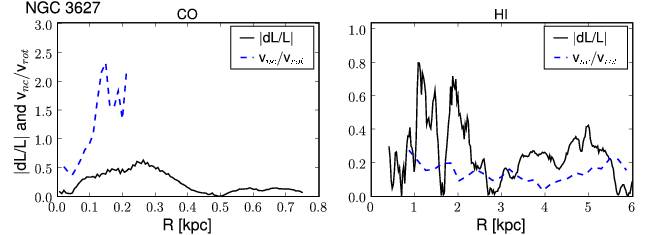}
\includegraphics[scale=0.62]{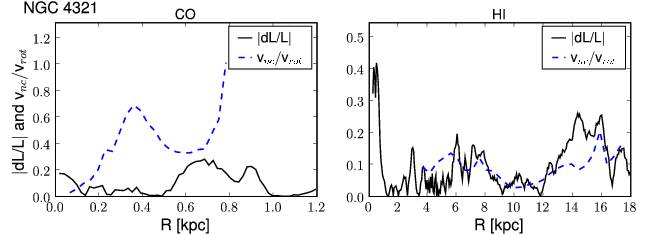}
\end{center}
\caption{\footnotesize Comparison between the absolute value of angular momentum transport per rotation $\vert$d$L/L \vert$ derived from our torque analysis (see \S \ref{subsec:torque_results}) with the fraction of non-circular to rotation velocity $v_{nc}/v_{rot}$ from the harmonic decomposition of the velocity field (see \S \ref{subsec:kin_model}). As non-circular motions are very likely dominated by elliptical streaming this comparison can be interpreted as elliptical streaming versus inflow/outflow as a function of radius. Values of $v_{nc}/v_{rot}$ with error bars larger than three (two) times the median of the dataset are not displayed for HI (CO).}
\label{fig_comp}
\end{figure}
\begin{figure}[ht]
\begin{center}
\includegraphics[scale=0.62]{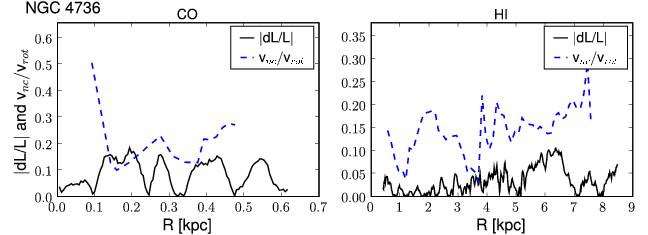}
\includegraphics[scale=0.62]{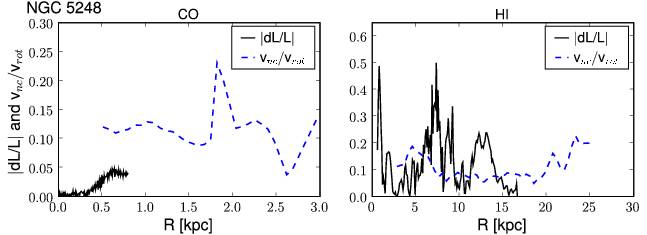}
\includegraphics[scale=0.62]{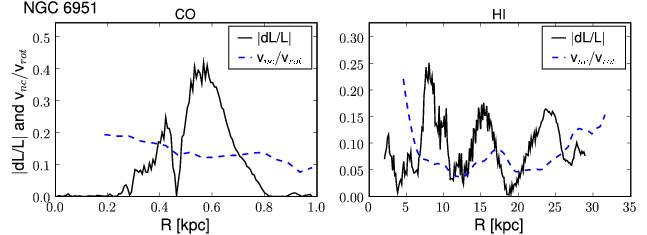}
\includegraphics[scale=0.62]{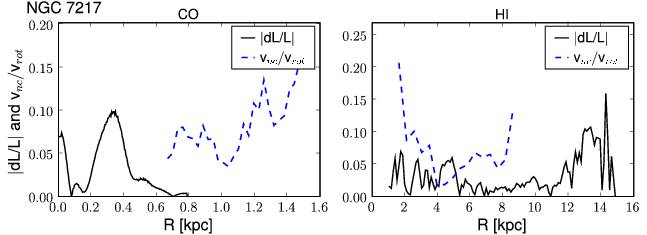}
\end{center}
\figurenum{\ref{fig_comp}}
\caption{(Continued)}
\end{figure}

\end{document}